\documentclass[review, 12pt]{elsarticle}
\usepackage{natbib}
\setcitestyle{square,numbers,sort&compress,comma}
\usepackage{amsmath}
\usepackage{amssymb}
\usepackage{caption}
\usepackage{graphicx}
\usepackage{latexsym}
\usepackage{times}
\usepackage{subfigure}
\usepackage{mlmath}

\journal{Combustion and Flame}

\usepackage{xcolor}  
\newcommand{\xzq}[1]{{\color{black} #1}}
\newcommand{\zxc}[1]{{\color{black} #1}}
\newcommand{\zth}[1]{{\color{black} #1}}
\newcommand{\rev}[1]{{\color{black} #1}}
\begin{document}
\begin{frontmatter}

\title{A multi-scale sampling method for accurate and robust deep neural network to predict combustion chemical kinetics}

\author[f,g,h]{Tianhan Zhang\corref{cor1}}
\author[a,b,f]{Yuxiao Yi}
\author[d,f]{Yifan Xu}
\author[d,f]{Zhi X. Chen}
\author[a,b,c]{Yaoyu Zhang}
\author[e,f]{Weinan E}
\author[a,b,f]{Zhi-Qin John Xu\corref{cor1}}

\address[a]{Institute of Natural Sciences, School of Mathematical Sciences, Shanghai Jiao Tong University, Shanghai, 200240, China}
\address[b]{MOE-LSC and Qing Yuan Research Institute, Shanghai Jiao Tong University, Shanghai, 200240, China}
\address[c]{Shanghai Center for Brain Science and Brain-Inspired Technology, Shanghai, 200240, China}
\address[d]{State Key Laboratory of Turbulence and Complex Systems, Aeronautics and Astronautics, College of Engineering, Peking University, Beijing, 100871, China}
\address[e]{School of Mathematical Sciences, Peking University, Beijing, 100871, China}
\address[f]{AI for Science Institute, Beijing, 100080, China}
\address[g]{Department of Mechanics and Aerospace Engineering, SUSTech, Shenzhen, 518055, China}
\address[h]{Department of Mechanical and Aerospace Engineering, Princeton University, NJ, 08540, US}
\cortext[cor1]{Corresponding authors.}

% -------------------------------------------------------------------- %
% -------------------------------------------------------------------- %
% -------------------------------------------------------------------- %

\begin{abstract} 
Machine learning has long been considered a black box for predicting combustion chemical kinetics due to the extremely large number of parameters and the lack of evaluation standards and reproducibility. The current work aims to understand two basic questions regarding the deep neural network (DNN) method: what data the DNN needs and how general the DNN method can be. Sampling and preprocessing determine the DNN training dataset, and further affect DNN prediction ability. The current work proposes using Box-Cox transformation (BCT) to preprocess the combustion data. In addition, this work compares different sampling methods with or without preprocessing, including the Monte Carlo method, manifold sampling, generative neural network method (cycle-GAN), and newly-proposed multi-scale sampling. Our results reveal that the DNN trained by the manifold data can capture the chemical kinetics in limited configurations but cannot remain robust toward perturbation, which is inevitable for the DNN coupled with the flow field. The Monte Carlo and cycle-GAN samplings can cover a wider phase space but fail to capture small-scale intermediate species, producing poor prediction results. A three-hidden-layer DNN, based on the multi-scale method without specific flame simulation data, allows predicting chemical kinetics in various scenarios and being stable during the temporal evolutions. This single DNN is readily implemented with several CFD codes and validated in various combustors, including (1). zero-dimensional autoignition, (2). one-dimensional freely propagating flame, (3). two-dimensional jet flame with triple-flame structure, and (4). three-dimensional turbulent lifted flames. The ignition delay time, laminar flame speed, lifted flame height, and contours of physical quantities demonstrate the satisfying accuracy and generalization ability of the pre-trained DNN. The Fortran and Python versions of DNN and example codes are attached in the supplementary for reproducibility, which can also be found on the https://github.com/tianhanz/DNN-Models-for-Chemical-Kinetics. 
\end{abstract}


\end{frontmatter}
\date{}
% -------------------------------------------------------------------- %
% -------------------------------------------------------------------- %
% -------------------------------------------------------------------- %

\clearpage

\section{Introduction} 

There is an increasing need to develop novel combustion techniques for demands like carbon-neutral goals and energy sustainability. Numerical simulation is a powerful tool to advance scientific discovery and assist industrial production. Chemists have developed many detailed combustion chemistry mechanisms to consider real fuel effects. However, it is a long-standing issue that simulations involving the direct integration (DI) of detailed chemistry are computationally expensive. Thus, developing numerical methods to accelerate simulations with detailed chemistry is a central topic within the combustion community \cite{Lu2009b}.

Machine learning has been introduced in the combustion area to accelerate computation since the 1990s. Christo et al. \cite{Christo1996, Christo1996a} were the first to combine the joint PDF method with artificial neural networks (ANNs) to predict chemical kinetics in turbulent jet diffusion flames. They compared global peak values and averaged radial profiles to demonstrate the reasonable accuracy of the ANN-PDF method. Blasco et al. \cite{Blasco1998} pointed out that it is necessary to evaluate the continuous evolution performance of an ANN in a homogeneous mixture before applying it to a flow system. Therefore, they investigated both the single and evolutionary prediction error of the ANN in a zero-dimensional system. Later they proposed a self-organizing map (SOM) approach to partition the thermochemical space into several subdomains where a group of ANNs can target each subdomain respectively \cite{Blasco2000}. The combination of the SOM and PDF-ANN achieved much higher accuracy in the temporal evolution of a partially stirred reactor (PaSR). Chen et al. \cite{Chen2000} emphasized the difficulty of sampling the entire chemical kinetics space and explored an innovative sampling strategy by the {\it in situ} adaptive tabulation (ISAT). It was shown that the ANN-ISAT could handle similar conditions under which the ISAT was built. Kempf et al. \cite{Kempf2005} trained ANNs to tabulate steady flamelets and tested them in piloted flame. A similar effort was made by Ihme et al. \cite{Ihme2009} to explore optimal ANN structures. To extend the ANN training data from steady-state solutions to unsteady simulations, Sen and Menon \cite{Sen2009} sampled data from direct numerical simulations (DNS) of laminar flame vortex interactions (FVI). They then formed a LEM-LES-ANN scheme without reducing the ANN input size using a progress variable or mixture fraction. They extended the sampling method to stand-alone LEM simulations and demonstrated its stronger generalization ability in a range of configurations \cite{Sen2010}. Chatzopoulos and Rigopoulos \cite{Chatzopoulos2013} suggested a tabulation method with Rate-Controlled Constrained Equilibrium (RCCE) and ANN for alleviating the pressure to parametrize higher-dimensional chemical space. They tested ANNs coupled with the RANS-PDF simulation for the DLR jet flames. Franke et al. \cite{Franke2017} further included flamelets of a broader range of strain rates in ANN training set to simulate Sydney flame L, which features local extinctions and re-ignitions. Wan et al. \cite{Wan2020} proposed an end-to-end deep neural network (DNN) in a non-premixed oxy-flame DNS trained on stochastic micro-mixing data. The end-to-end DNN could directly predict burning rates without a PDF model and achieved satisfactory accuracy in the flame statistics. Ding et al. \cite{Ding2021} examined an innovative hybrid sampling strategy to train the ANN, including both flamelet and random data, aiming for a stronger generalization. Their ANN was tested on flamelets, one-dimensional premixed flame, and Sandia flame series. Chi et al. \cite{Chi2021} put forward an on-the-fly training scheme to accelerate DNS, and the ANN does not require preliminary knowledge and can learn in an in situ manner. More recently, Nakazawa \cite{Nakazawa2022} proposed a species-independent DNN structure so that the trained models are not limited to a single configuration. \zth{Another typical data-driven method is principal component analysis (PCA), which can find the low dimensional subspace of the reacting system to reduce variable numbers in the simulation, including the PC-score approach \cite{Sutherland2009, Malik2021} and KPCA approach \cite{Mirgolbabaei2014}. The PCA can further improve computation efficiency by adaptive chemical kinetics in different thermochemical domains \cite{DAlessio2020}.}

It is readily seen that the generalization, accuracy, and efficiency of ANN/DNN models are at the core of previous studies. An ideal ANN/DNN model should be able to handle a wide range of temperature, pressure, and equivalence ratio conditions in various flow and flame configurations with or without turbulence. However, there are still major barriers to developing such a model. The first one is the generalization ability, which heavily relies on training data selection. In past works, sampling was performed on random data \cite{Christo1996,Christo1996a,Blasco1998, Blasco2000}, ISAT \cite{Chen2000}, flamelet \cite{Kempf2005,Ihme2009,Chatzopoulos2013,Franke2017}, DNS \cite{Sen2009, Nakazawa2022}, stand-alone LEM simulation \cite{Sen2010}, stochastic micro-mixing \cite{Wan2020}, randomly perturbed flamelet data \cite{Ding2021} and on-the-fly simulation \cite{Chi2021}. The neural networks aforementioned were validated through various challenging cases and demonstrated good agreement. It is nevertheless still unclear whether they can be applied to a wider range of conditions. If a given configuration needs fine-tuning of a specific ANN, the usefulness of such methods in combustion research will be very limited.

Therefore, thorough validations are required to demonstrate the applicability of ANN/DNN for realistic problems. On the one hand, an accurate single-step prediction may not reflect the model's robustness in a temporal evolution and against flow perturbations. On the other hand, turbulent combustion cases are expensive and difficult to analyze. The combined effort is thus necessary to provide a convincing validation but is so far scarce. 

With the above objectives, in this work, we propose a multi-scale sampling method with the preprocessing method of the Box-Cox transformation to collect multi-scale combustion data. The sampling method is applied to study a wide thermochemical phase space
%: temperature from 800 K to 3100 K, pressure 0.5 atm to 2 atm, 
with no explicit constraints on equivalence ratio, flow field, or turbulence. Common techniques such as Monte Carlo sampling, manifold sampling, and cycle-GAN method \cite{zhu2017unpaired}, are compared with the multi-scale sampling with or without preprocessing. A three-hidden-layer DNN is off-line trained based on the multi-scale sampling data. The error evaluations of the DNN are performed by single-step prediction, continuous evolution prediction, and evolution with perturbation prediction. The model is then implemented into CFD codes and tested for a broad range of configurations: (1) zero-dimensional homogeneous ignition simulations with ignition delay time comparison; (2) one-dimensional transient premixed flame with laminar flame speed comparison; (3) two-dimensional laminar jet flame with triple-flame structure; (4) three-dimensional turbulent lifted jet flame. 

The paper is constructed as follows: first, the sampling methodology and preprocessing are proposed and compared with other sampling methods. Second, deep neural network structures are investigated to guarantee sufficient efficiency and robustness. Third, systematic validations are performed. Finally, the conclusions are drawn. 
%The resulting DNN is attached in the Supplementary. Example codes are also provided.

\section{Methodology} 
\subsection{Multi-scale combustion data}

Generally speaking, there are two types of multi-scales in combustion chemical kinetics: the multi-scale concentration and reaction rate distributions of different species and species at different reaction stages. In other words, the data in different dimensions in thermochemical phase space have divergent orders of magnitude and characteristic time scales during the chemistry system evolution. The inherent multi-scale features lead to non-linear and stiff combustion chemistry. Meanwhile, it is a major challenge for data-driven methods to capture the fundamental physics from the multi-scale dataset instead of fitting data by brutal force. 

The multi-scale phenomena have been widely discussed in the combustion area. However, it is still vital to understand the multi-scale features of combustion from the perspective of data. A typical example of the difficulty associated with the multi-scale mass fraction is plasma-assisted combustion \cite{Ju2015a}. In contrast to fuels and oxidizers, OH concentration and its temporal change are small. The deep neural network tends to focus on major changes such as fuel and oxidizer while ignoring small OH radical changes. But adding OH radicals with a mass fraction of $10^{-5}$ in the preheat zone can dramatically accelerate the combustion process \cite{Zhang2021}. It is far more effective than adding more fuel or oxidizer, even if their mass fractions are several orders of magnitude larger than the radicals. The small concentration change of OH results in a huge difference in chemical kinetics, implying that a large weight for OH concentration should be learned in a deep neural network. The large change of the output induced by a small change of the input is also termed high frequency. The frequency principle has shown that deep neural network is difficult to learn high-frequency information \cite{xu_training_2018,xu2019frequency}. In addition, simply increasing the weight of OH cannot solve the problem. Since OH radical concentration is much higher in the burned zone than in the premixed zone, adding OH radical can hardly change the burned gas chemical states. In other words, OH should have a small weight for deep neural network prediction in the burned gas. The multi-scale distributions of different species and species at different times require the deep neural network to distill the raw data and capture the underlying physics. 

In physics, small quantities are as important as other quantities, but data science needs explicit preprocessing of the raw data to address this issue. The result of a lack of preprocessing is that different neural networks can only deal with chemical kinetics in different phase space subdomains. Additionally, it is more difficult for the neural networks without preprocessing to capture intermediate species. In the current work, the Box-Cox transformation is adopted to overcome the difficulty of the multiple magnitude scales. It is elaborated in section 2.2. 

\subsection{Box-Cox transformation}\label{sec:bct}
The Box-Cox transformation (BCT), which was originally proposed by Box and Cox (1964) \cite{Box1964}, is defined as
\begin{equation}
	\mathcal{F}(x)=\begin{cases}
		\displaystyle\frac{x^{\lambda}-1}{\lambda}~~~~, & \lambda\neq0 \\
		\log x,~~~~                                     & \lambda=0
	\end{cases}
	\label{eq:BCT}
\end{equation}
where $x$ is the variable and $\lambda$ is a hyper-parameter. 

\begin{figure}
	\centering
	\ifx\mycmd\undefined
	\includegraphics[width=0.6\textwidth]{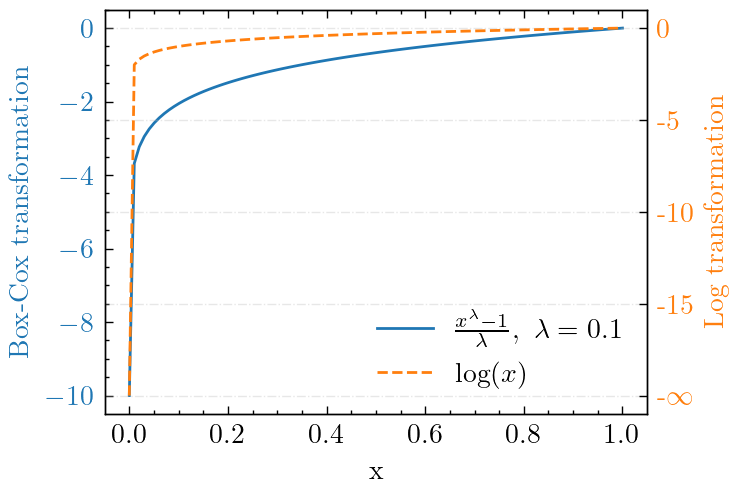}
	\fi
	\caption{$\log(x)$ and BCT with $\lambda=0.1$.
	}
	\label{fig:bct}
\end{figure}

Compared with the log transformation, BCT not only has the advantage of representing multi-scale data by  $\mathcal{O}(1)$ quantity,
but also avoids the singularity of the log transformation when the data is approaching zero.
For the mass fraction of the chemical species within $[0,1]$, the BCT maps $[0,1]$ to $[-1/\lambda,0]$.
Meanwhile, BCT maps the low-order quantity $10^{-k}(k\geq 2)$ to $\sim\mathcal{O}(1)$ with appropriate $\lambda$. In this work, we use $\lambda=0.1$. As an example, we show log-transformation and BCT with $\lambda=0.1$ in Fig \ref{fig:bct}. \rev{Note that in the current work, we adopt BCT only for species mass fractions since mass fractions have typical multi-scale distributions. Other physical quantities such as temperature and pressure do not need BCT to pre-preprocess. After BCT of the mass fractions, we use the Z-score Normalization for all dimensions of the training data, including temperature, pressure, and mass fractions, which is a widely-used normalization method in the machine learning area.}
\subsection{Deep neural network}\label{sec:dnn}
%$\mathcal{F}(Y_i(t))_{i=1,\cdots,n}$ is strange
We use $\vx(t)=\{T(t),P(t),\mathcal{F}(Y_i(t))_{i=1,\cdots,n}\}$ to denote the temperature, pressure, and mass fraction of each species at evolution time $t$. 
The DNN takes input $\vx$ to produce an output $\vu_{\vtheta}(x)$, which aims to predict the change $\vu^{*}(t):=\vx(t+\Delta t)- \vx(t)$. \zth{Note that only mass fractions are obtained from DNN prediction. The temperature and density are calculated according to enthalpy and mass conservation. More specifically, the density and enthalpy in each grid are updated by calculating interface fluxes. After the DNN predicts the local mass fraction changes, the temperature is calculated through implicit iterations given the enthalpy, density, and updated mass fractions.}
The training dataset $D = \{\vx_{i},\vu^{*}_{i}\}_{i=1}^{N}$ is Z-score normalized (subtracted by mean and divided by standard deviation) for both input and output, where $N$ is the sample size.

We use the following $L$-layer neural network to learn dataset $D$,
\begin{align}
    \vu_{\vtheta}(\vx)& = \vW^{[L-1]} \sigma\circ(\mW^{[L-2]}\sigma\circ(\cdots ( \nonumber\\
    &\mW^{[1]} \sigma\circ(\mW^{[0]} \vx + \vb^{[0]} ) + \vb^{[1]} )\cdots) \nonumber\\
    &+\vb^{[L-2]})+\vb^{[L-1]},\nonumber
\end{align}
where $\mW^{[l]} \in \sR^{m_{l+1}\times m_{l}}$, $\vb^{[l]}=\sR^{m_{l+1}}$, $m_0=m_{L}=n+2$, ``$\circ$'' means entry-wise operation, $\sigma$ is Gaussian Error Linear Unit (GELU) \cite{Hendrycks2016}. We denote the set of parameters by $\vtheta$. The loss function is mean absolute error ($L_1$ loss) and is
defined as
\begin{align}
	Loss=\frac{1}{N}\frac{1}{n+2}\sum_{i=1}^{N}\left\Vert\vu^{*}_i-\vu_{\vtheta}(\vx_i)\right\Vert_{L_1}
	\label{eq:loss}
\end{align}
% where $N$ is the number of samples.
% , $\vu_{\vtheta,i}:=\vu_{\vtheta(\vx_i)}$.

\subsection{Failure of vanilla sampling methods}
\subsubsection{Manifold sampling with and without BCT and its limitation}
Given initial condition space of temperature $T$,
pressure $P$ and equivalence ratio $\Phi$, ensembles of the thermochemical states during the chemical system evolution form
a smooth manifold \cite{Maas1992}. 
\zth{It is worth noticing that the manifold sampling borrows the concept of the intrinsic manifold but does not reduce the dimension of the chemical kinetics. The sampling is conducted during the transient combustion simulations to collect thermochemical states.}
Similar to previous works \cite{Zhanga}, for hydrogen/air mixture, 5000 initial conditions are randomly sampled from
$T\in\left[1000K, 1500K\right]$, $\Phi\in\left[0.5, 3\right]$ and  $P\in\left[0.5atm, 1.5atm\right]$. The chemistry mechanism contains eight species and 16 reversible reactions \cite{Evans1980}. \xzq{Each case is simulated in Cantera with a constant time step size $10^{-8}$ s until the mixture reaches chemical equilibrium. The thermochemical state $\vx(t)$ pairs with its temporal changes $\vu^{*}(t) ={\vx(t+\Delta t)-\vx(t)}, \Delta t = 10^{-6} s$ to form a data point. It is worth noticing that not all thermochemical states are included in the dataset due to the extremely large number of data points. Instead, the thermochemical states with higher temporal changing rates are more likely to be selected.}
In total, 1,700,000 samples are obtained from the 5,000 ignition cases for the training and testing.
Unless otherwise specified, we use $D_m$ to represent manifold dataset where the subscript \textit{m} denotes \textit{manifold}. 

\begin{figure}[!h]
	\centering
	\ifx\mycmd\undefined
	\includegraphics[width=\textwidth]{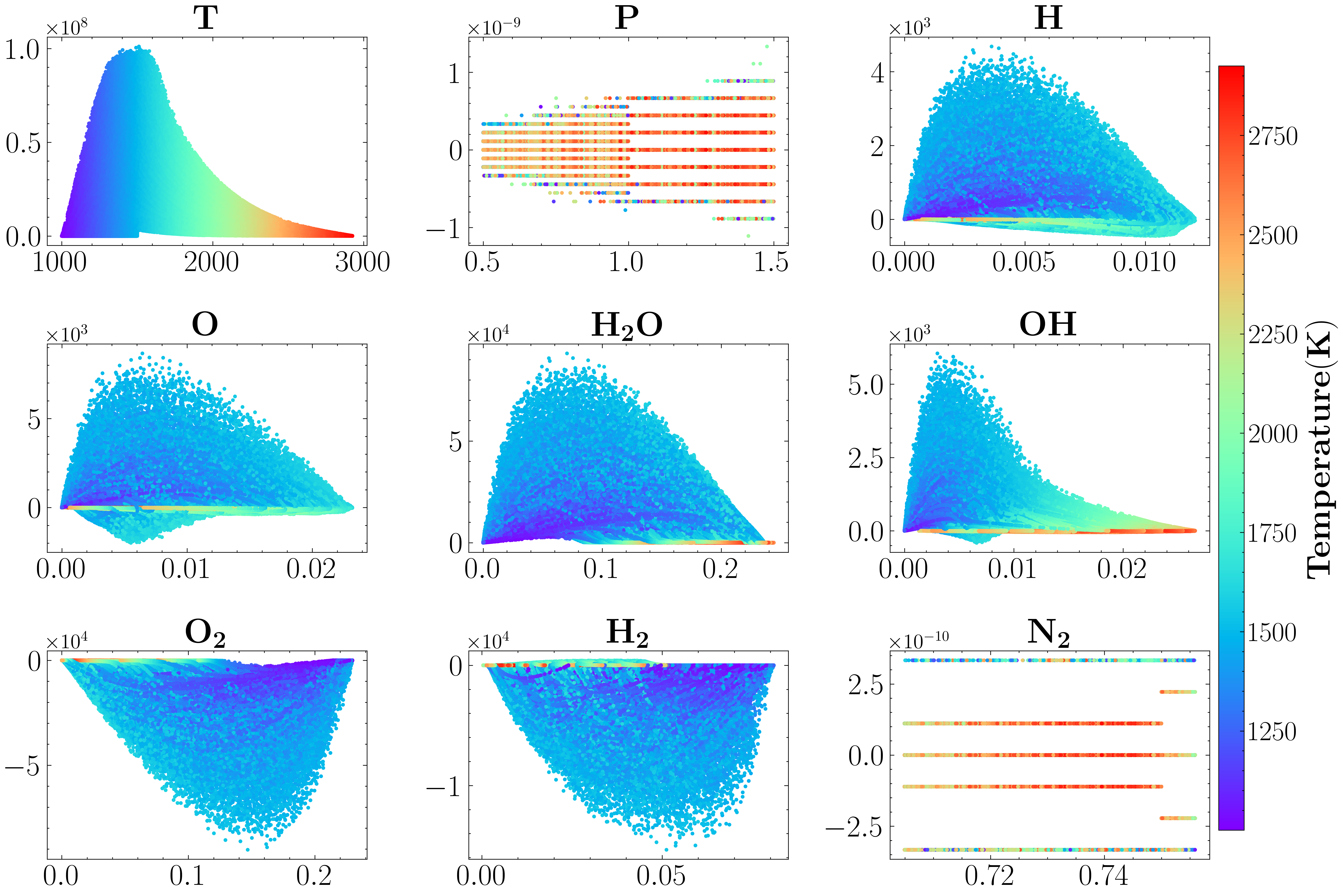}
	\fi
	\caption{Phase diagram of the manifold dataset, $T_0\in\left[1000K, 1500K\right]$, $\Phi\in\left[0.5, 3\right]$ and  $P\in\left[0.5atm, 1.5atm\right]$. Each sub-figure illustrates the temporal change rate (ordinate) against the value (abscissa) for temperature, pressure and the concentrations of species, respectively. Color indicates temperature.
	}
	\label{fig:phase_manifold}
\end{figure}

Figure \ref{fig:phase_manifold} shows the collected data by manifold sampling. The x-axis is the value of the thermochemical state, and the y-axis is its temporal gradient. The phase diagram shows the multi-scale features in combustion. It is seen that there exist several orders of magnitude differences for values and their temporal gradients in different thermochemical dimensions. In addition, the peaks of the temporal gradients are around $T = 1500 K$, while the temporal gradients are close to zero at the initialization and burned stages.

\begin{figure}
	\centering
	\ifx\mycmd\undefined
	\subfigure[without BCT]
	{
	    \includegraphics[width=0.45\textwidth]{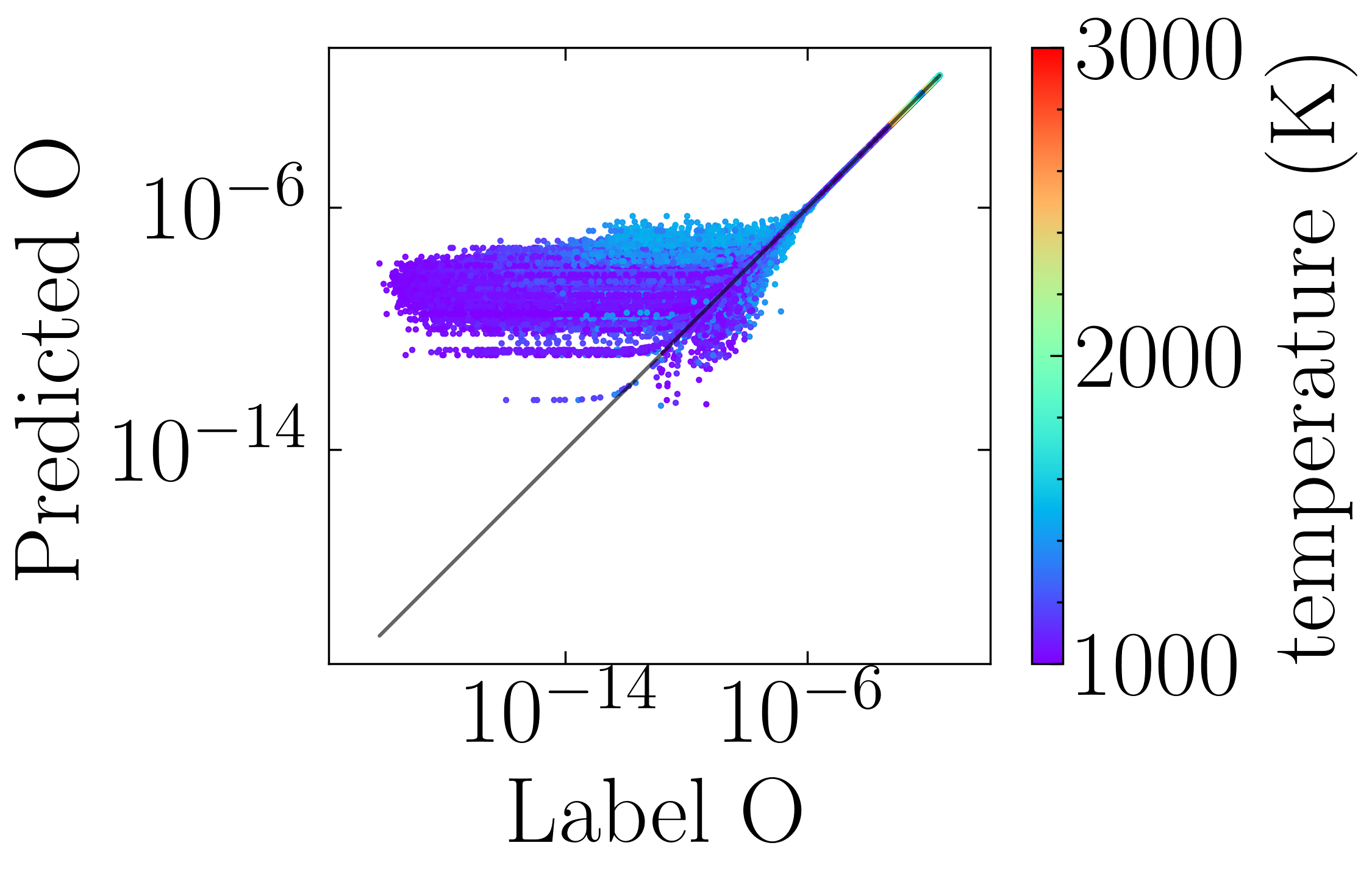}
	}
	\subfigure[with BCT]
	{ 
	    \includegraphics[width=0.45\textwidth]{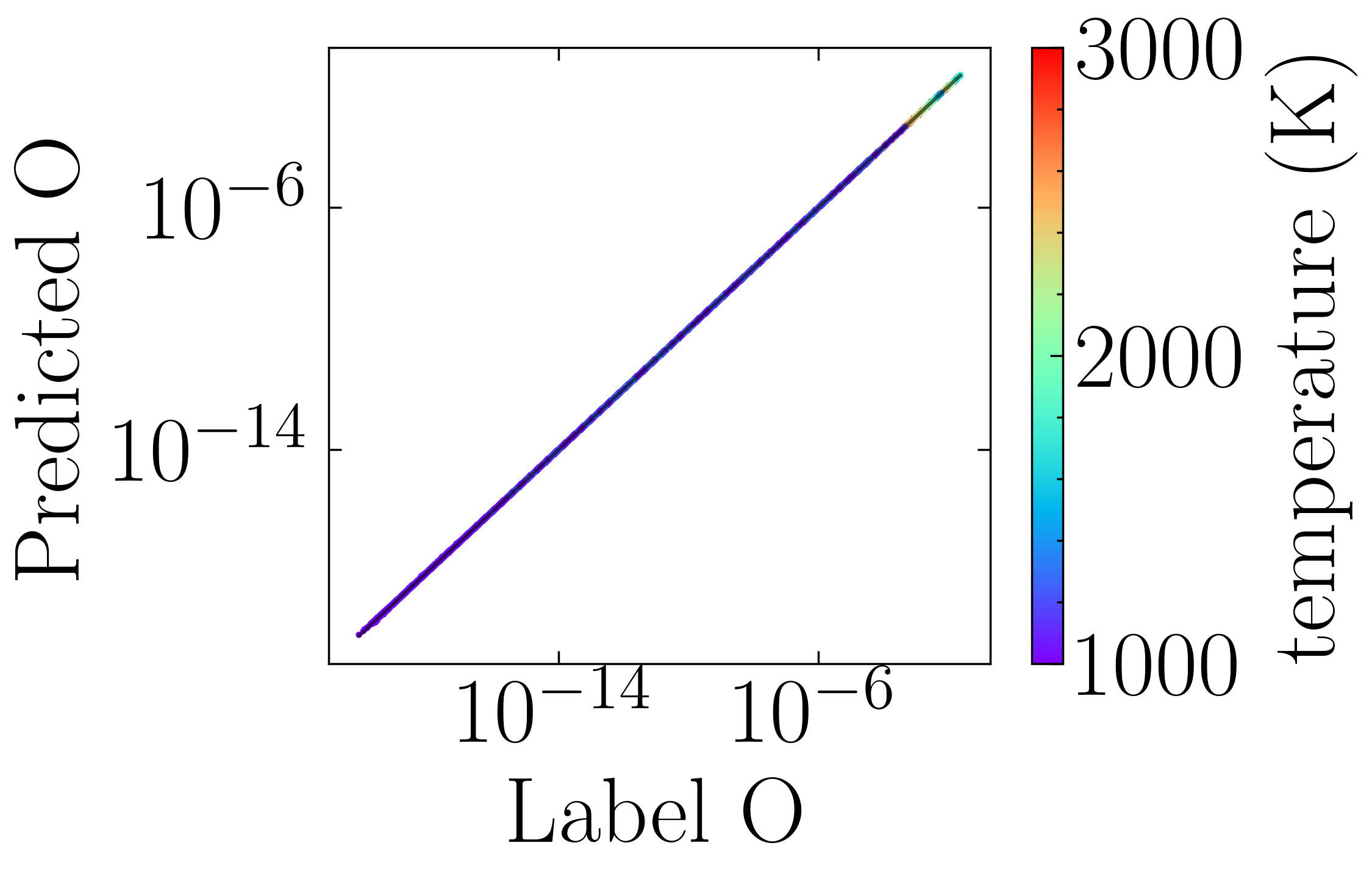}
	}
	\subfigure[continuous evolution]
	{
	    \includegraphics[width=0.4\textwidth]{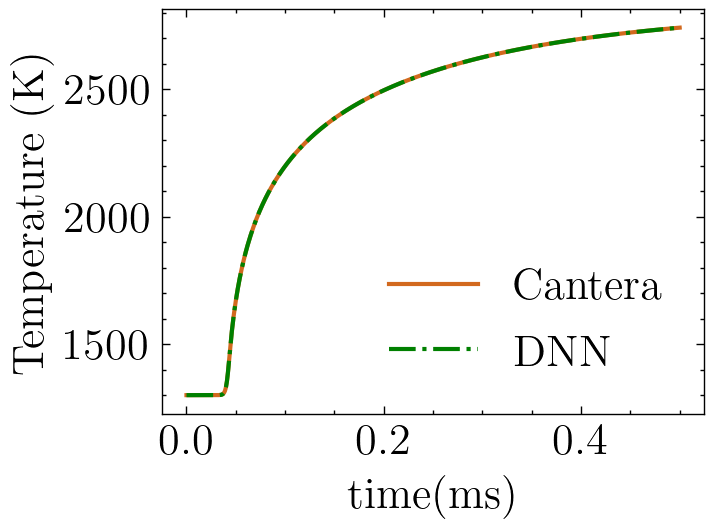}
	}
	\subfigure[test on perturbation]
	{ 
	    \includegraphics[width=0.45\textwidth]{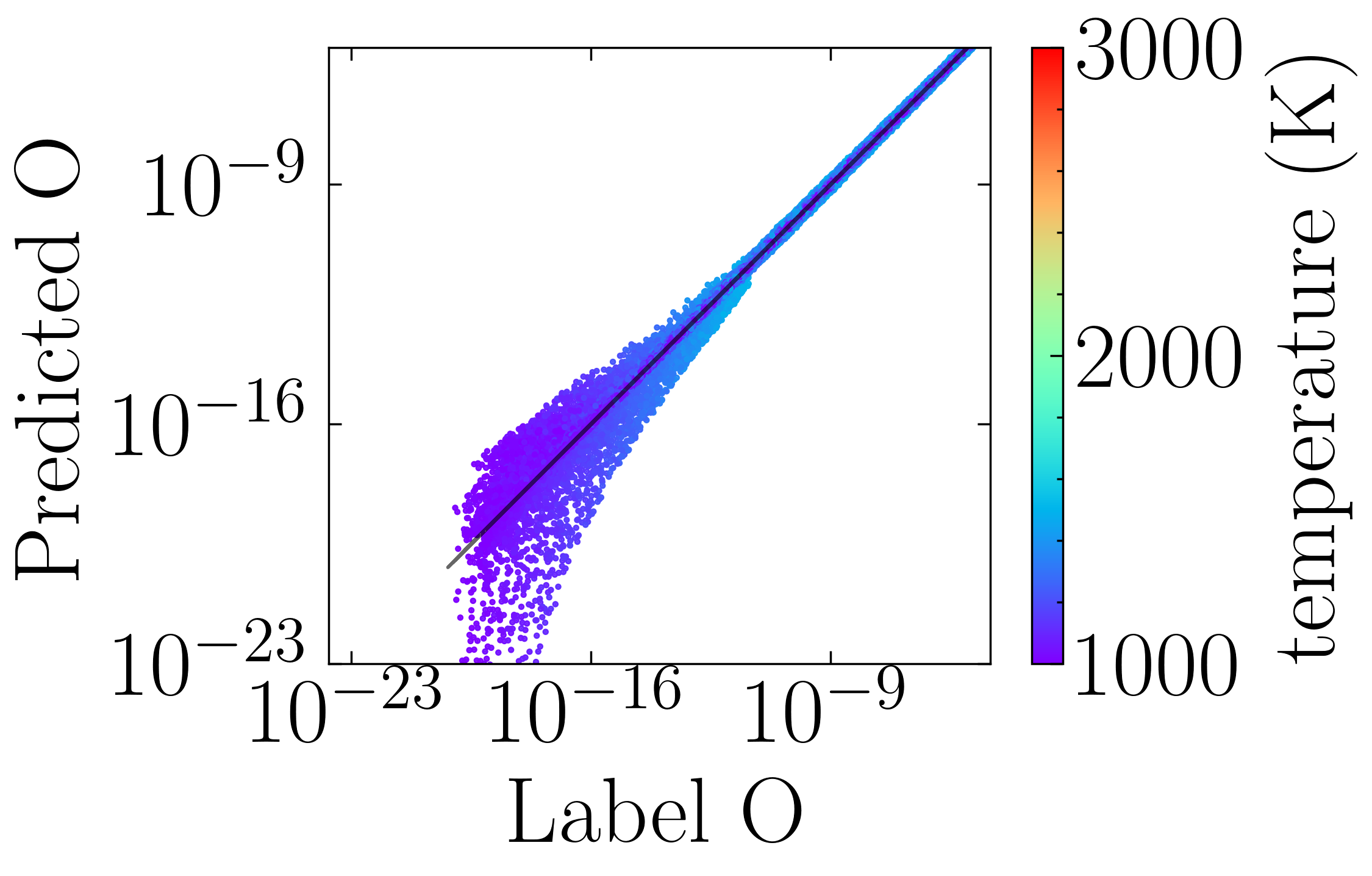}
	}
	\fi
	\caption{Prediction of $O$ production rate by Cantera (abscissa) and DNN (ordinate) trained by manifold sampling on (a) manifold samples without BCT; (b) manifold samples with BCT; (c) continuous evolution on the manifold; (d) perturbed samples with perturbation coefficient $\alpha$=5. The color of each dot indicates the temperature of the sample.
	}
	\label{fig:mansam}
\end{figure}

To further explore the current DNN’s applicability, the DNN prediction test on a perturbed dataset is performed. Here, we perturb the samples as follows. For each state vector in $D_m$, we randomly add or subtract $\alpha\%$ of the species mass fraction, that is,
\begin{align*}
	\hat{Y}_i=Y_i\times (1 \pm \alpha\%),
\end{align*}
where $Y_i$ denotes mass fraction of species except $N_2$. The mass fraction of $N_2$ is determined by other species concentrations since their summation is unity.
The labels are generated by Cantera. As is shown in Figure \ref{fig:mansam}, the performance of DNN are excellent on predicting manifold data in b and c, but completely fails on perturbed samples in d. Since the perturbations can be quite common when the DNN is coupled with the flow field, the poor performance on the perturbed dataset indicates that a wide range of sampling is necessary for training a robust DNN.

\subsubsection{Monte Carlo method and the challenge from small-scale radical distribution}

\begin{figure}[!h]
	\centering
	\ifx\mycmd\undefined
	\includegraphics[width=\textwidth]{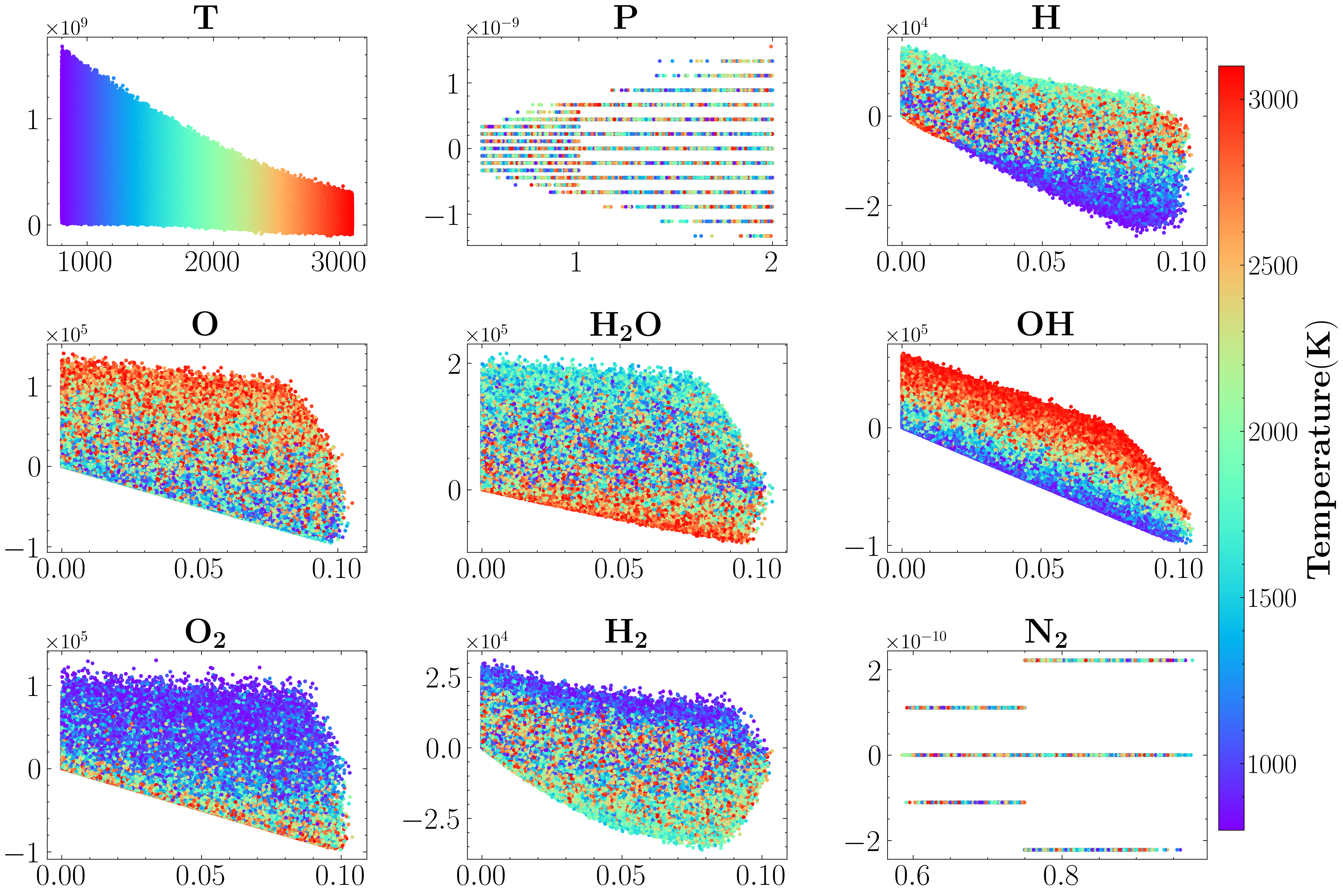}
	\fi
	\caption{Phase diagram of dataset of MC sampling method, $T\in[800K, 3100K]$, $P\in[0.5atm, 2atm]$, $Y(N_2)\in[0.6, 0.8]$ and $Y(*)\in[0, 1]$, $*$ denotes the rest species except $N_2$. Each sub-figure illustrates the temporal change rate (ordinate) against the value (abscissa) for temperature, pressure and the concentrations of species, respectively. Color indicates temperature.
	}
	\label{fig:phase_MC}
\end{figure}

It is natural to utilize Monte Carlo (MC) method to generate a wide range of random samples.
In this study, the range of working conditions are set as follows: $T\in[800K, 3100K]$, $P\in[0.5atm, 2atm]$, $Y(N_2)\in[0.6, 0.8]$ and $Y(*)\in[0, 1]$ where the marker $*$ denotes the rest species except $N_2$. 
\xzq{Each dimension, including temperature, pressure, and mass fractions, is sampled randomly with uniform probability in the given range. The species mass fractions are normalized to guarantee summation equal unity.} The current MC sampling collects 6,000,000 samples and covers the manifold discussed in the last section. Figure \ref{fig:phase_MC} shows a broader distribution of the Monte Carlo sampled phase space compared with the manifold sampling results. The stochastic colors determined by the data point temperature also indicate the randomness of the sampled data. However, the DNN trained by this larger dataset fails to fit $D_{m}$, which is the data on the manifold. 

\xzq{
To uncover the reason why the Monte Carlo method failed to sample desired data, a data distribution similar to the manifold dataset $D_{m}$ is reproduced by cycle-GAN, a type of generative adversarial network specializing in capturing data distribution patterns. That is to say, cycle-GAN is a middle-way choice balancing the Monte Carlo sampling and the manifold sampling. 
On the one hand, cycle-GAN data distribution is expected to be close to the manifold sampling. On the other hand, it is important for cycle-GAN to have randomness to some extent. This constrained randomness provides a good comparison with the Monte Carlo method to analyze data sampling strategy and quality. 
The temperature and the pressure are sampled in the same way as the Monte-Carlo method while the mass fractions of all species are sampled through a cycle-GAN. The cycle-GAN is trained to build up the mapping between a simple distribution, for example, normal Gaussian distribution, and the target distribution. In the current work, the cycle-GAN consists of two GANs. Each GAN consists of two components, a generative network, and a discriminative network. For the first GAN, the generative network's input is the data sampled from a seven-dimensional normal Gaussian distribution. The output is a seven-dimensional vector that represents mass fractions. The generative network is fully-connected with hidden layer sizes of 200-800-1000-800-400-200. The data generated by the GAN network is labeled as 0 while the data sampled from the manifold is labeled as 1. Data sampled from the generative network and manifold method is used to train the discriminative network, which is a fully-connected network with hidden layer sizes of 200-600-200. The discriminative network aims to classify the manifold data and the generated data, while the generative network is trained to generate more challenging data for the discriminative network to classify, in other words, to increase the error of the discriminative network so that the discriminative network becomes more accurate. The second GAN maps the data generated by the first GAN back to the normal Gaussian distribution in a similar way. These two GANs are trained iteratively until the generated data distribution of the first GAN is indistinguishable from the manifold data so that the generative network can produce a dataset with a similar distribution pattern as the manifold sampling.
In the end, a dataset $D_{GAN}$ is generated by the cycle-GAN with 6,000,000 samples.
}

\begin{figure}
	\centering
	\ifx\mycmd\undefined
	\subfigure[]
	{ 
	    \includegraphics[width=0.45\textwidth]{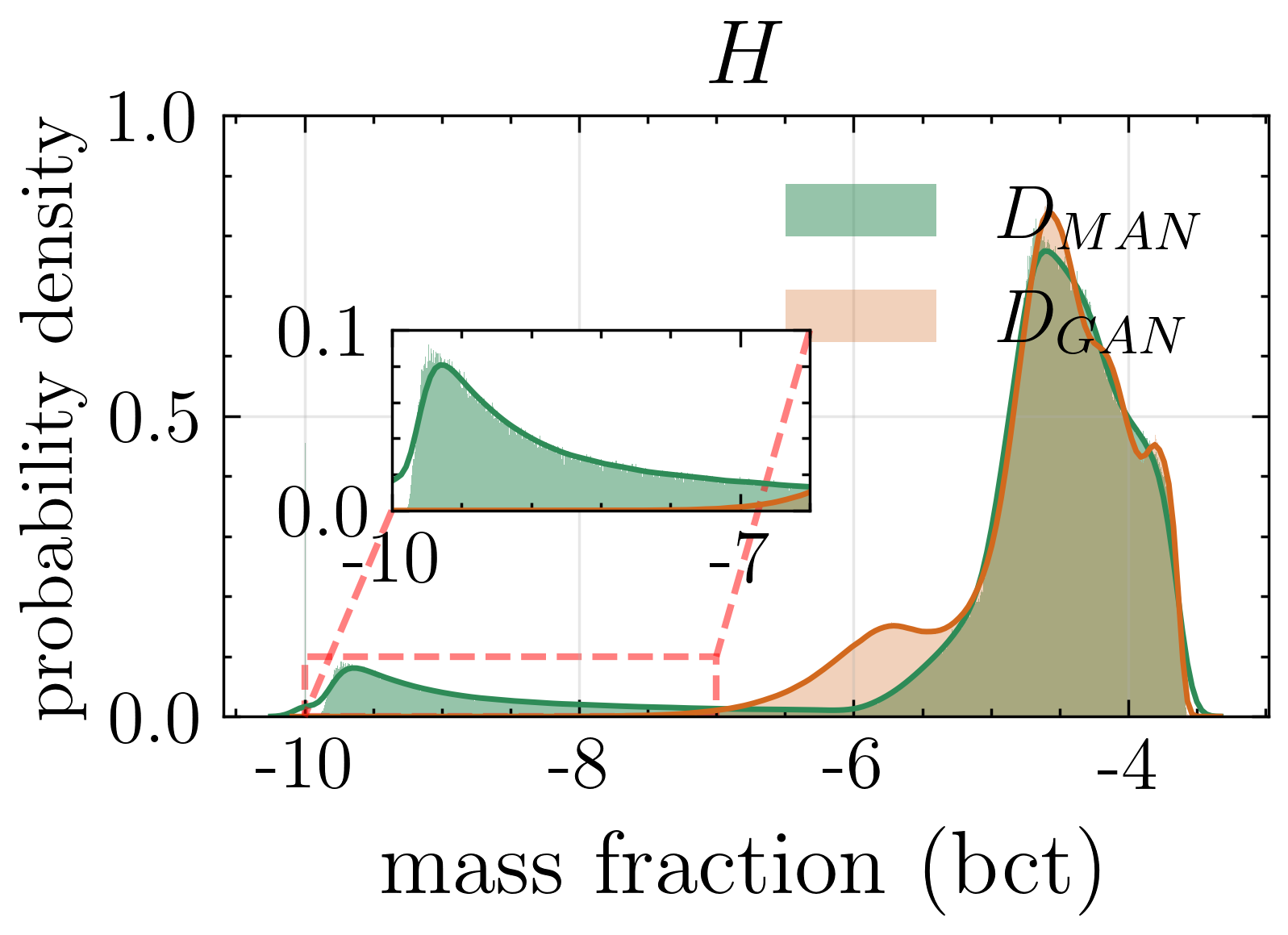}
	}
	\subfigure[]
	{ 
	    \includegraphics[width=0.45\textwidth]{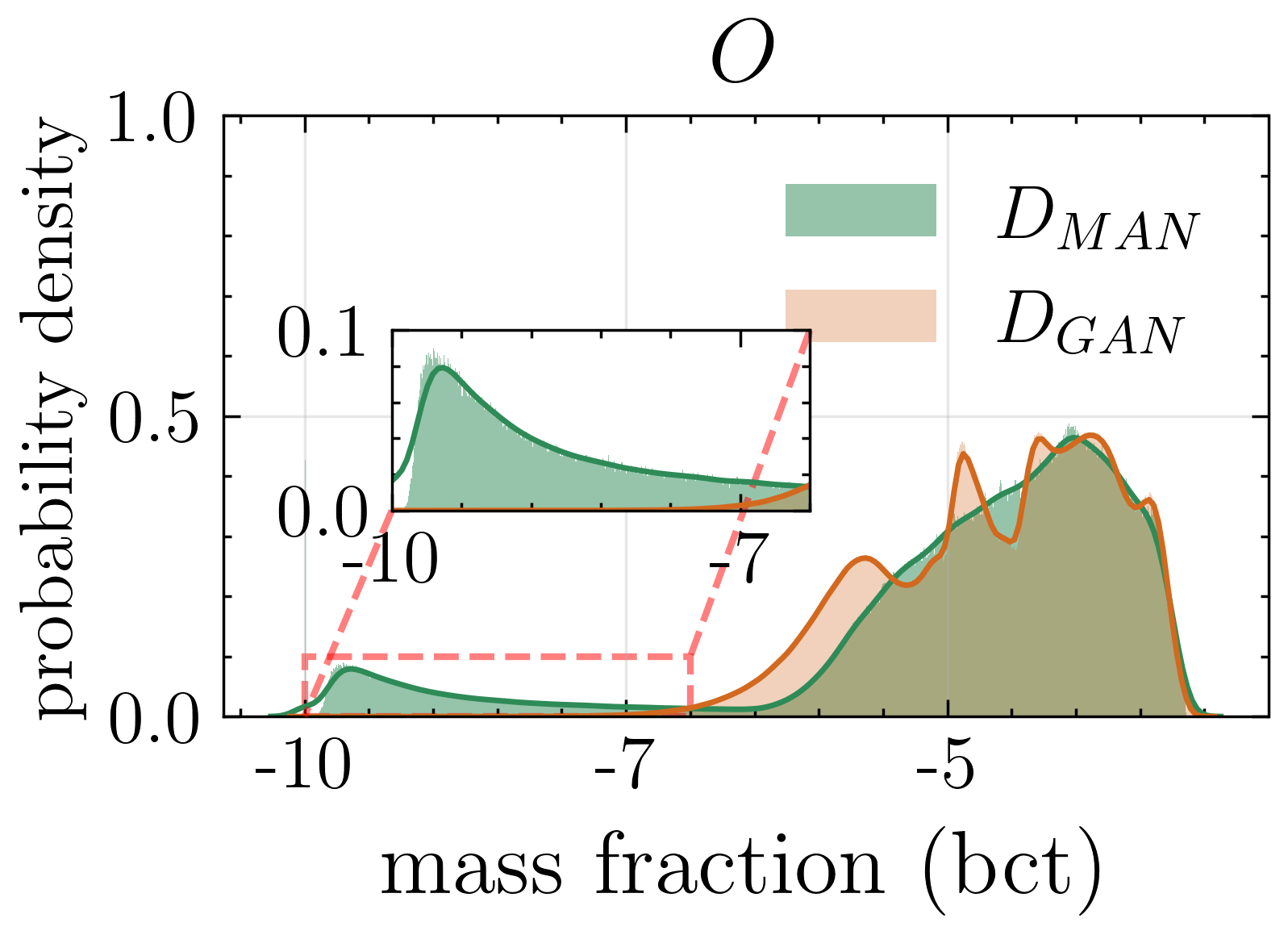}
	}
	\fi
	\caption{The distributions of radicals (a) $H$ and (b) $O$ by manifold sampling and cycle-GAN sampling are compared, respectively. The subset shows that the distribution of $D_{GAN}$ does not cover small scales data in the manifold.
	}
	\label{fig:ganmanifold}
\end{figure}
\begin{figure}
	\centering
	\ifx\mycmd\undefined
	\subfigure[GAN on $D_s$]
	{
		\includegraphics[width=0.45\textwidth]{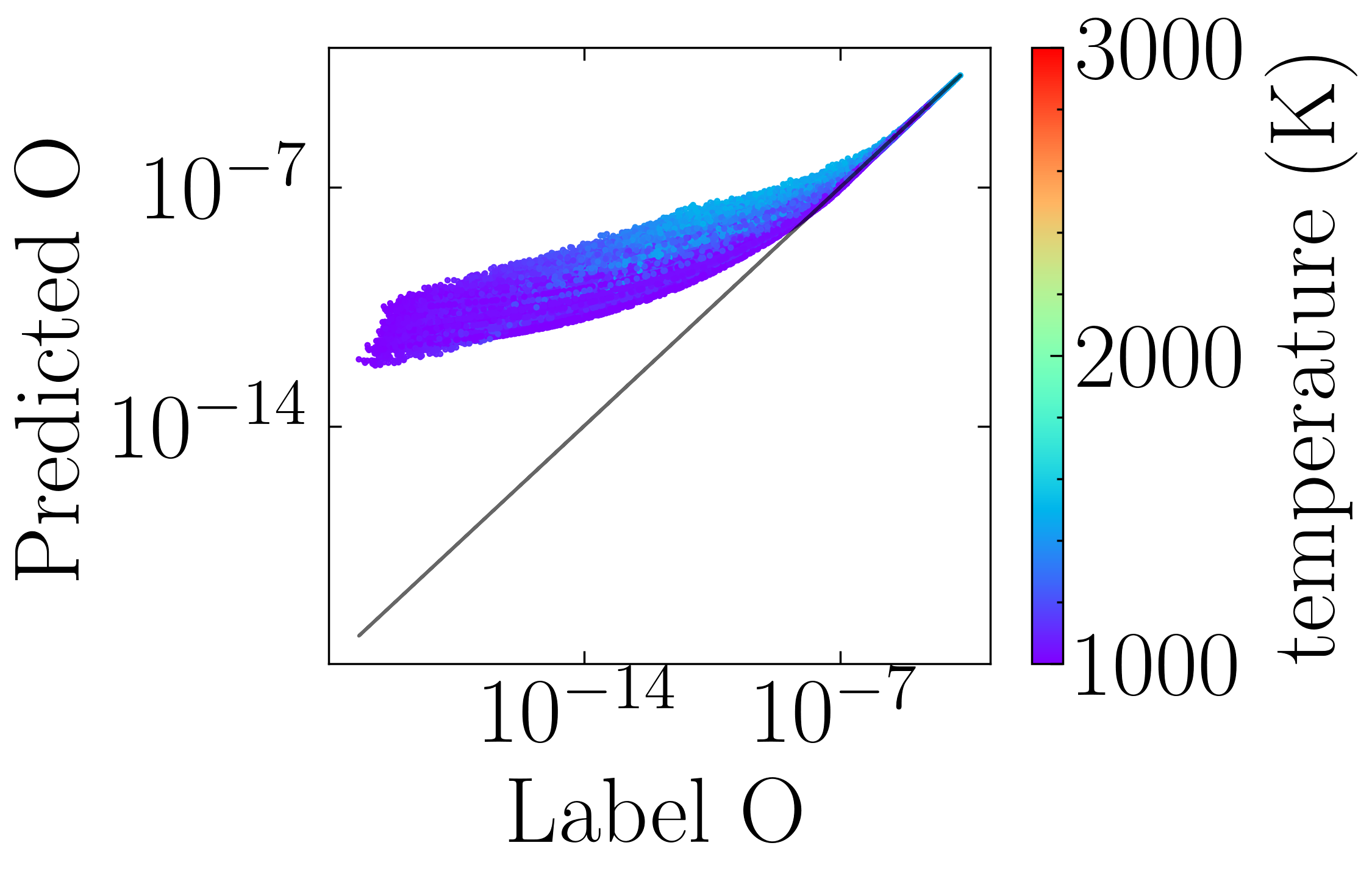}
	}
	\subfigure[GAN on $D_c$]
	{
		\includegraphics[width=0.45\textwidth]{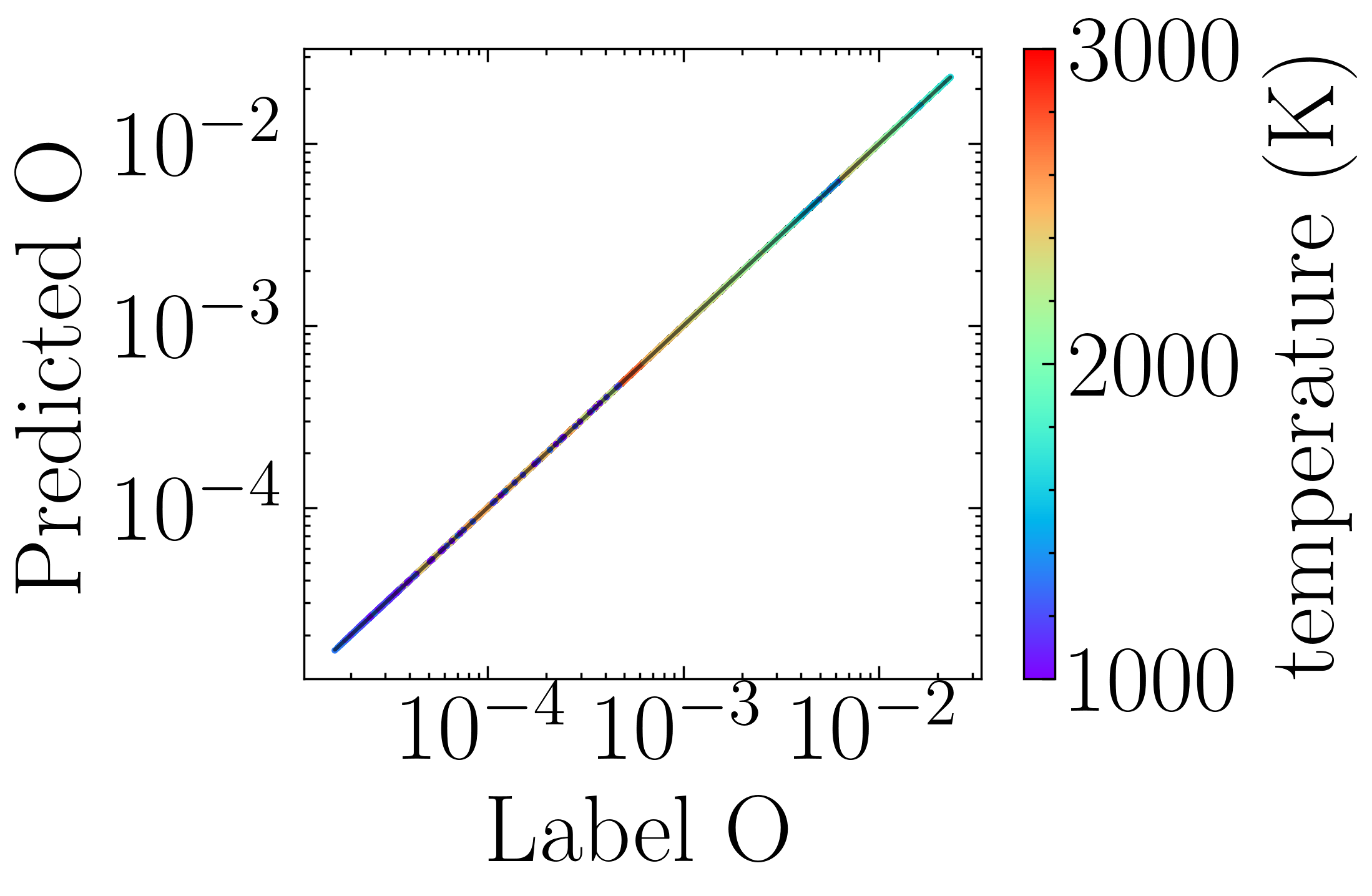}
	}
	\subfigure[MC on $D_s$]
	{
		\includegraphics[width=0.45\textwidth]{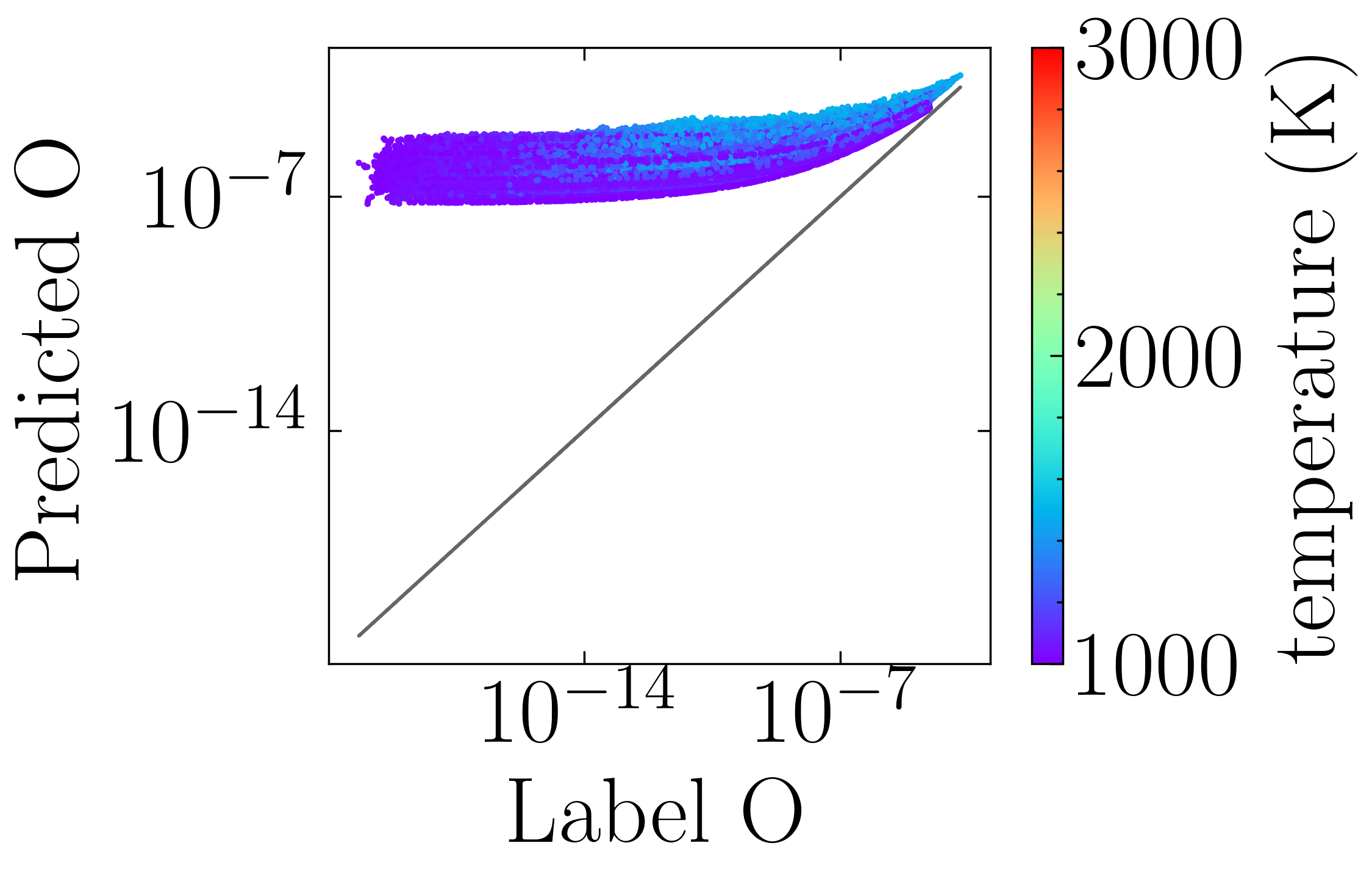}
	}
	\subfigure[MC on $D_c$]
	{
		\includegraphics[width=0.45\textwidth]{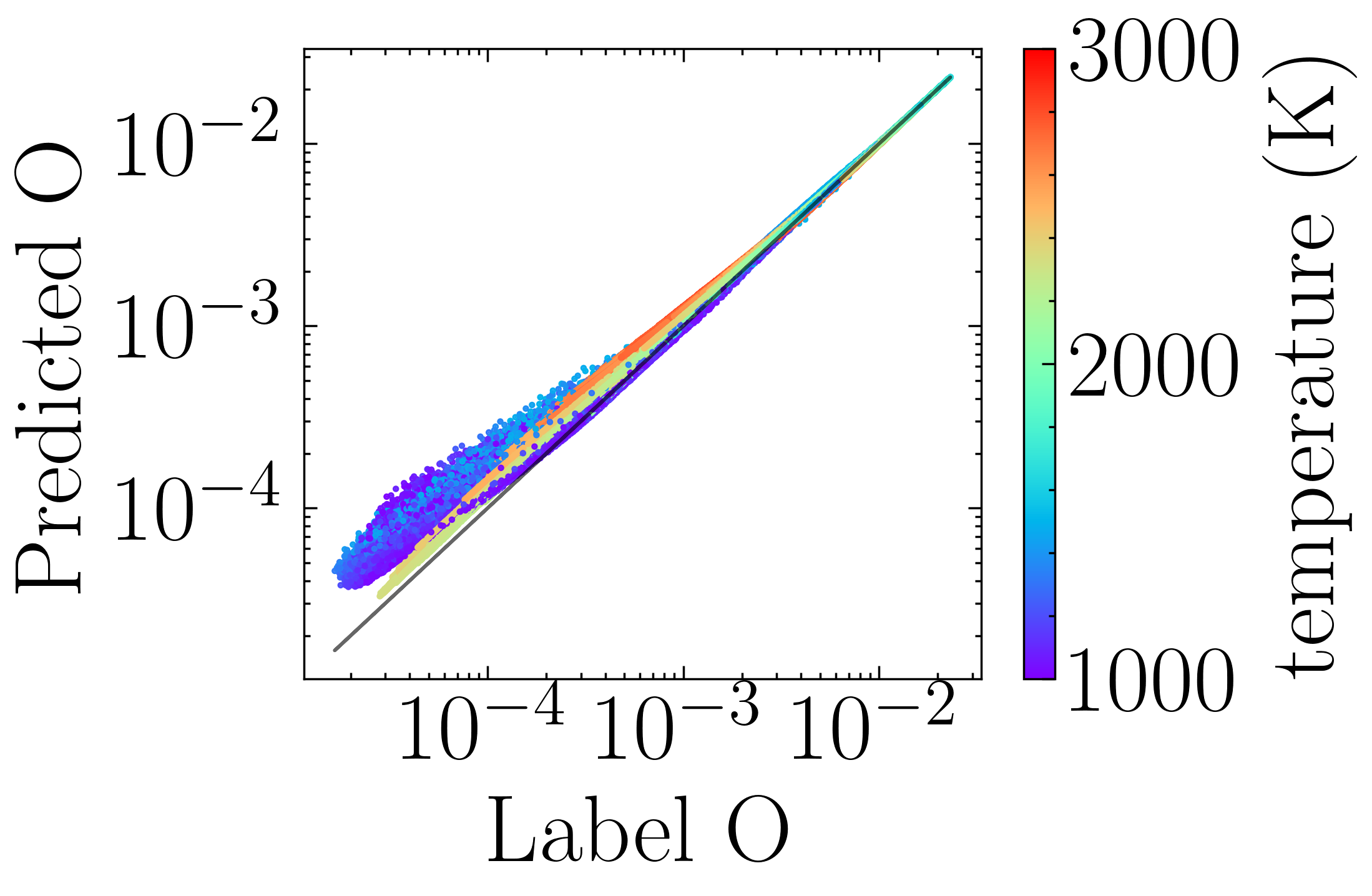}
	}
	\fi
	\caption{Prediction of $O$ mass fraction by Cantera (abscissa) and DNN (ordinate) trained by GAN/MC sampling method  on small-scale manifold data in (a,c), on large-scale manifold data in (b,d). 
	}
	\label{fig:gan}
\end{figure}

\zth{
Figure \ref{fig:ganmanifold} shows the distribution comparison of H and O mass fractions between the manifold data (green) and the cycle-GAN data (orange). Two distributions are similar in general.} It turns out that the major difference between the manifold data and cycle-GAN data is the small-scale radical distribution. On the one hand, cycle-GAN data distribution is consistent with the manifold data on large scales, i.e., the mass fraction above $10^{-5}$. However, two sampling methods generate distinct data distributions when radical mass fraction below $10^{-7}$. We then decompose the manifold dataset $D_m$ into two parts, i.e., $D_m=D_s\cup D_c$, where ‘s’ means ‘small’ and $D_s$ contains samples of which at least one radical mass fraction is smaller than ${10}^{-5}$, and ‘c’ means ‘complementary’ and $D_c$ contains the rest of $D_s$ data. \zth{
In other words, $D_m$ represents the data range that the three sampling methods (manifold sampling, Monte Carlo sampling, and cycle-GAN sampling) cover in common, while $D_s$ is mainly captured by the manifold sampling. Using $D_s$ and $D_c$ to evaluate DNNs trained on the Monte Carlo and cycle-GAN data, it can depict how sampled data distribution impact DNN training and predicting, especially the role of $D_s$. 
Figure \ref{fig:gan} shows that the DNN trained by $D_{GAN}$ or $D_{MC}$ performs well on $D_c$ but fails on $D_s$. It is clear the Monte Carlo and cycle-GAN methods are effective in sampling data in $D_c$. However, the main drawback is the lack of data points in $D_s$. As a result, the DNNs' divergent performances on $D_s$ and $D_c$ are within expectation since the DNN cannot predict well where the training data is not enough. The divergent performance of the two datasets indicates the importance of considering multi-scale features for a generic sampling method.
}

\subsection{Multi-scale sampling method}
\begin{figure}[!h]
    \centering
	\includegraphics[width=0.5\textwidth]{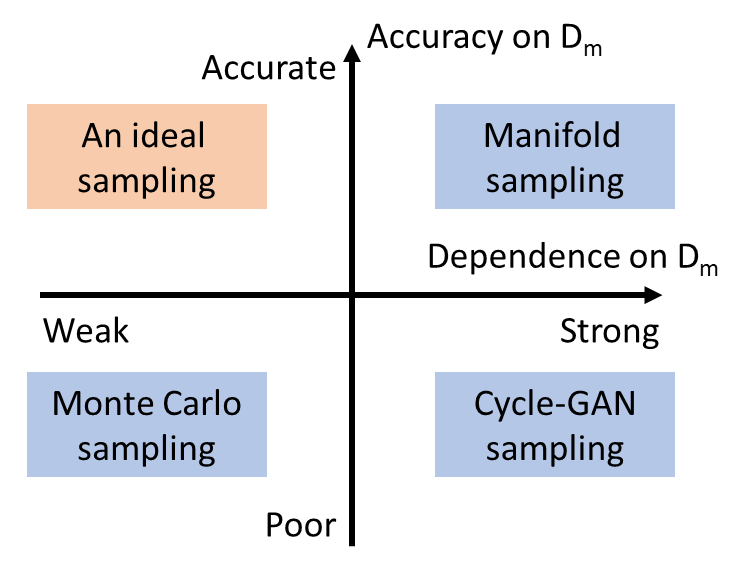}
	\caption{A comparison of different sampling methods. $D_m$ is the dataset by the manifold sampling.
	}
	\label{fig:comparison}
\end{figure}

Figure \ref{fig:comparison} summaries three sampling methods mentioned above. The sampling methods are characterized by their dependence and accuracy on $D_m$. The DNN trained on $D_m$ has the highest accuracy and a strong dependence on the manifold data. The main drawback is the limited applicability of the DNN away from the manifold. Ding et al. \cite{Ding2021} proposed a sampling method combining manifold data (flamelet data in their case) and random data. Nonetheless, it remains unclear to what extent the random data improves DNN’s performance in a general scenario, such as homogeneous autoignition. Besides, the manifold sampling requires additional computation costs to resolve priori flame structures or autoignition evolutions. The Monte Carlo sampling and cycle-GAN sampling cannot predict accurately on $D_s$. However, their failure enlightens a new sampling method capturing different scales of random data.

\begin{figure}[!h]
    \centering
    \ifx\mycmd\undefined
	\includegraphics[width=\textwidth]{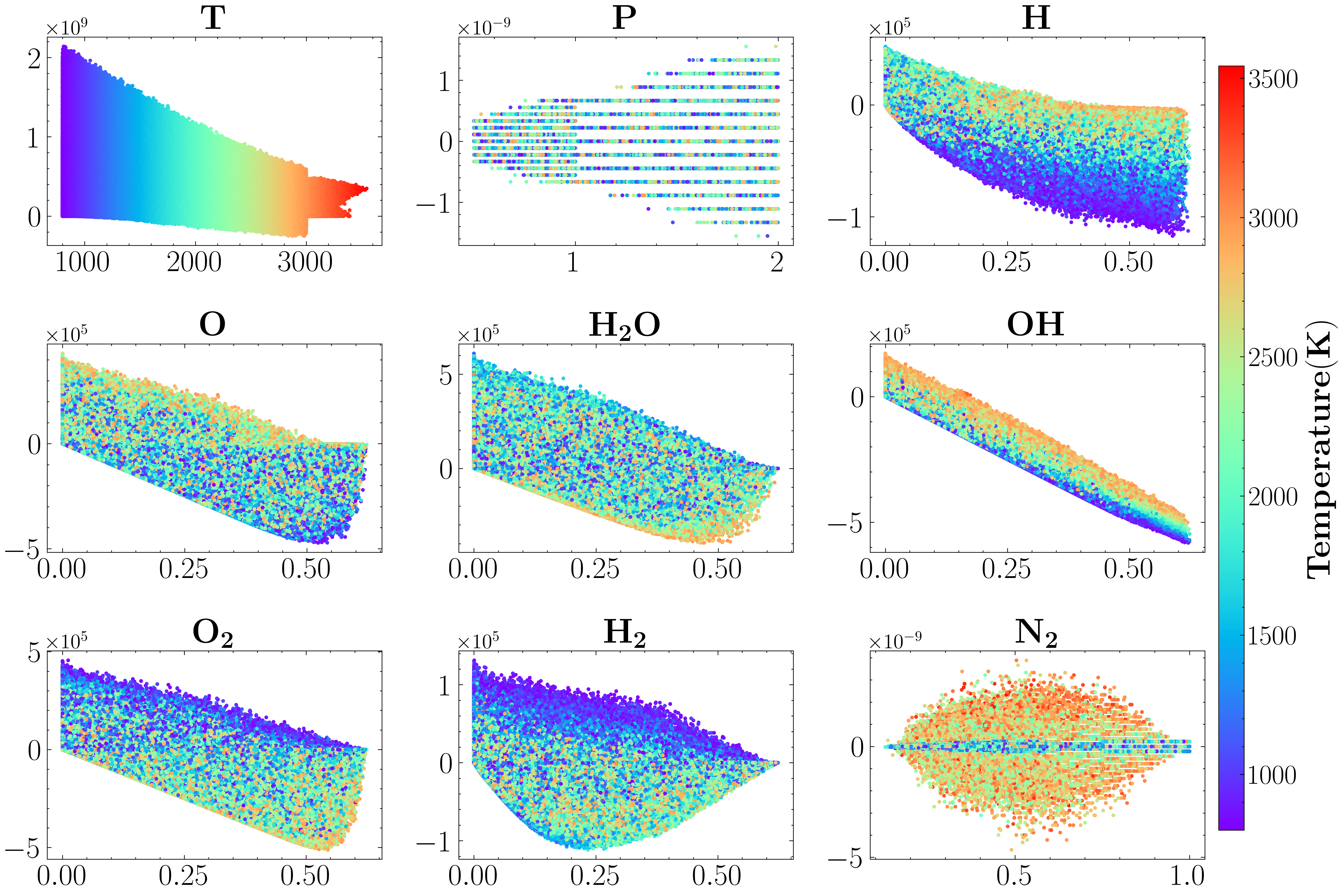}
	\fi
	\caption{Phase diagram of dataset of the multi-scale sampling method $T\in[800K, 3100K]$, $P\in[0.5atm, 2atm]$, and $Y(N_2)\in[0.2, 0.9]$. Each sub-figure illustrates the temporal change rate (ordinate) against the value (abscissa) for temperature, pressure and the concentrations of species, respectively. Color indicates temperature.
	}
	\label{fig:phase_MS}
\end{figure}

\begin{figure}
	\centering
	\ifx\mycmd\undefined
	\subfigure[test on manifold]
	{ 
	    \includegraphics[width=0.45\textwidth]{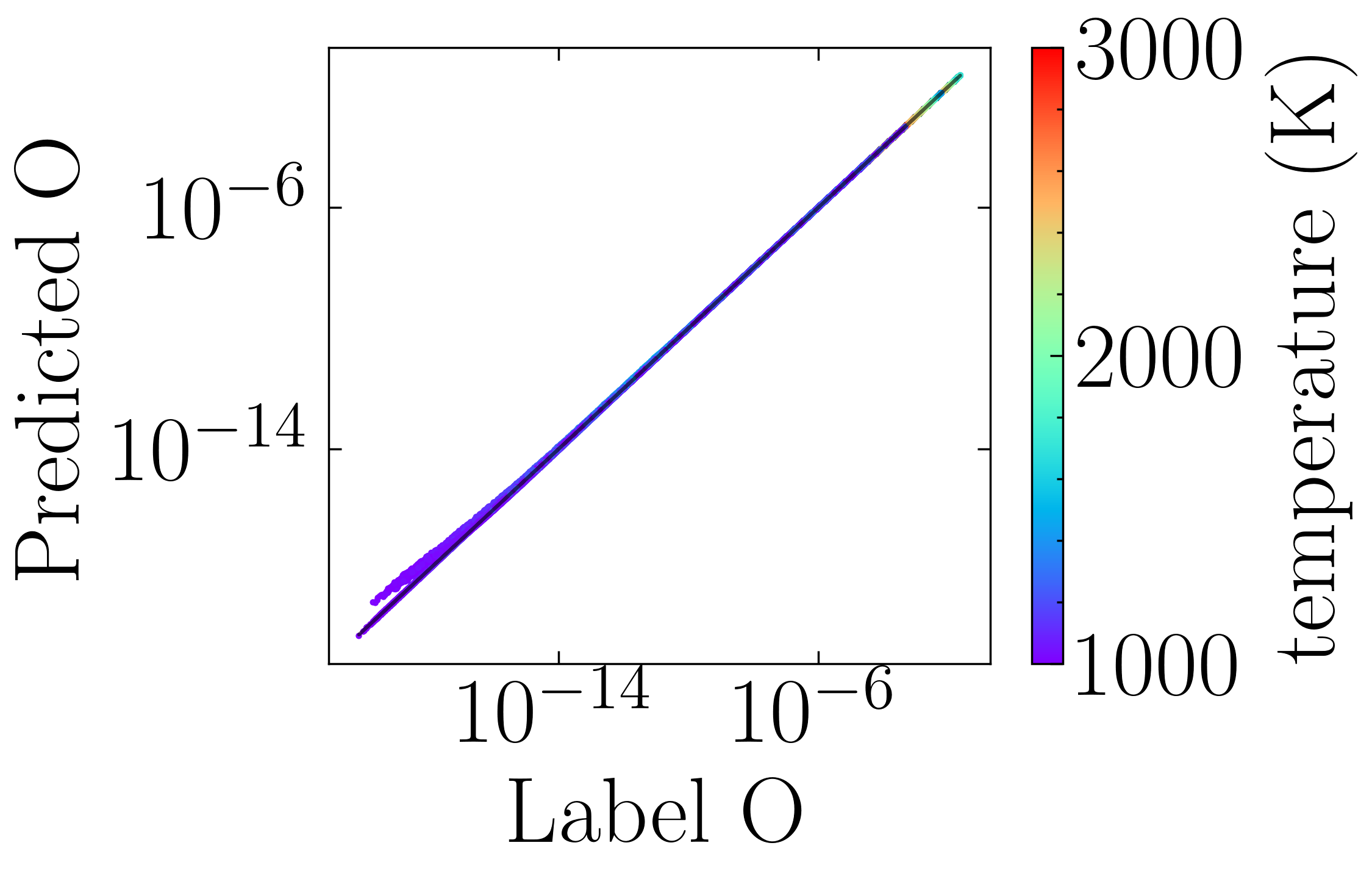}
	}
	\subfigure[test on perturbation]
	{ 
	    \includegraphics[width=0.4522\textwidth]{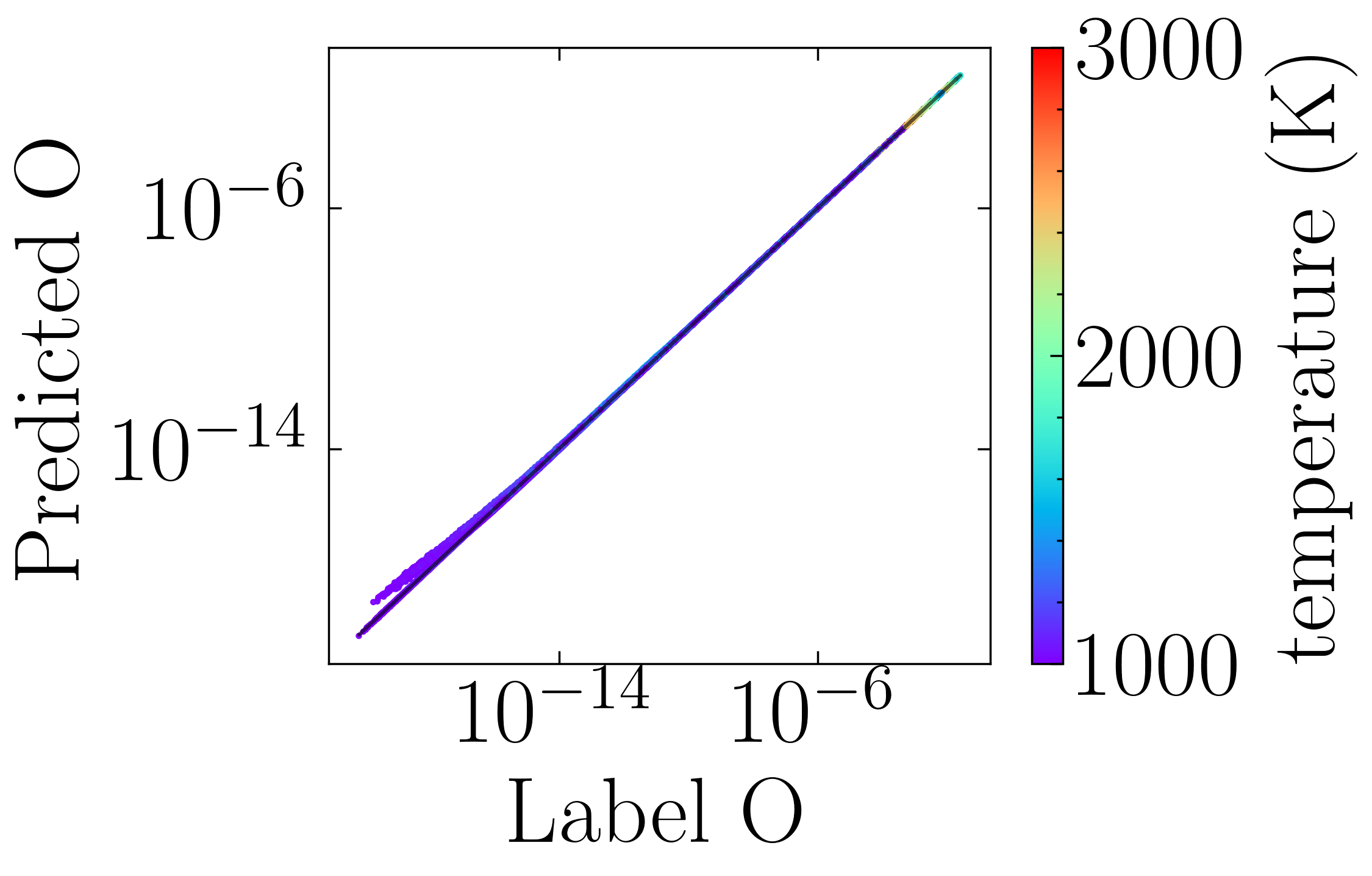}
	}
	\fi
	\caption{Prediction of $O$ by Cantera (abscissa) and DNN (ordinate) trained by multi-scale sampling method on manifold samples (a) and perturbed samples (b) with $\alpha$=5. The color  of each dot indicates the temperature of the sample.
	}
	\label{fig:MSsam}
\end{figure}

This work proposes a multi-scale sampling method targeting comprehensive coverage of multi-scale combustion data. Temperature and pressures are randomly sampled in a wide range: $T\in[800K, 3100K]$, $P\in[0.5atm, 2atm]$. The species are classified into two groups: major species and others. Here only $N_2$, fuel, and oxidizer are considered as major species and they are sampled randomly under log-scale in the following range: $Y(N_2) \in [0.2, 0.9], Y(H_2) \in [10^{-5}, 1], Y(O_2) \in [10^{-5},1]$. Other species are sampled randomly from $[10^{-k_i}, 1], k_i=1,…,25$ under log-scale. \zth{Note that the selection of major species is not based on the classical Major-minor species model. Instead, it chooses the species with relatively high concentrations throughout temporal evolutions. For example, $H_2O$ is the major product but is not considered a major species in the current model, because its mass fraction starts from zero and crosses over several magnitudes. From the data point of view, $H_2O$ is more similar to radicals.} 
180,000 samples are generated for each $k_i$ and a total 4,500,000 sample dataset is obtained. An important observation is that the DNN trained on the current 4,500,000 sample dataset can predict high-reaction-rate states well, but its performance is relatively worse on the thermochemical states from the burned gas. As a result, an additional dataset is added where $T\in [1800K, 3100K], P\in [0.5atm, 2atm], Y(N_2)\in [0.2, 0.9]$, and other species $Y(*) \in [10^{-4},1]$. 400,000 initial states are generated randomly in temperature, pressure, and log-scale species dimensions. For each initial state $\vx$, two temporally consecutive states are collected: $\vu^{*}(t)=\vx(t+\Delta t)- \vx(t)$ and $\vu^{*}(t+\Delta t)=\vx(t+2*\Delta t)- \vx(t+\Delta t)$. Finally, a complete dataset consisting of 5,300,000 samples is generated. The corresponding phase diagram is shown in Figure \ref{fig:phase_MS}. The current dataset’s range is the widest compared with manifold sampled data and Monte Carlo sampled data regarding the range of values or temporal gradients in each dimension. Figure \ref{fig:MSsam} shows that the DNN trained on the current multi-scale dataset is accurate on both manifold data in Figure \ref{fig:MSsam}a and perturbed data in Figure \ref{fig:MSsam}b. The results imply that the current sampling method helps achieve a smaller dataset on which a more accurate and generic DNN can be trained. There is no doubt that the multi-scale sampling method can be much more automatic and less ad hoc. Nevertheless, the main focus of such a sampling method will always be addressing the multi-scale data properly.


\section{Results and discussion}
In this section, a pre-trained DNN is validated by various benchmark tests. It is worth noticing that all the tests are performed using the same DNN. In the first subsection, the zero-dimensional autoignitions and one-dimensional transient premixed flames are simulated. These simulations can provide a detailed and quantitative comparison of ignition delay time and laminar flame speed between DNN and traditional methods, two of the most important quantities in combustion. A two-dimensional jet flame with a triple-flame structure is simulated in the second subsection using DNN and implicit Euler as ODE integrator. Since the triple-flame structure consists of a diffusion flame and premixed lean and rich flames, it is a good scenario to demonstrate the DNN performance across a wide range of local flow and thermochemical conditions. The third subsection further validates the DNN in turbulent flow scenarios, where a highly turbulent lifted flame experiment \cite{cheng1992} is simulated using the same DNN model deployed for the previous lower-dimensional and laminar cases. This test case serves as strong support for the generality and robustness of the proposed multi-scale sampling method-based DNN in complex flow configurations.

\subsection{Zero- and one-dimensional tests}

\begin{figure}[!h]
	\centering
	\ifx\mycmd\undefined
	\subfigure[]
	{ 
		\includegraphics[width=0.9\textwidth]{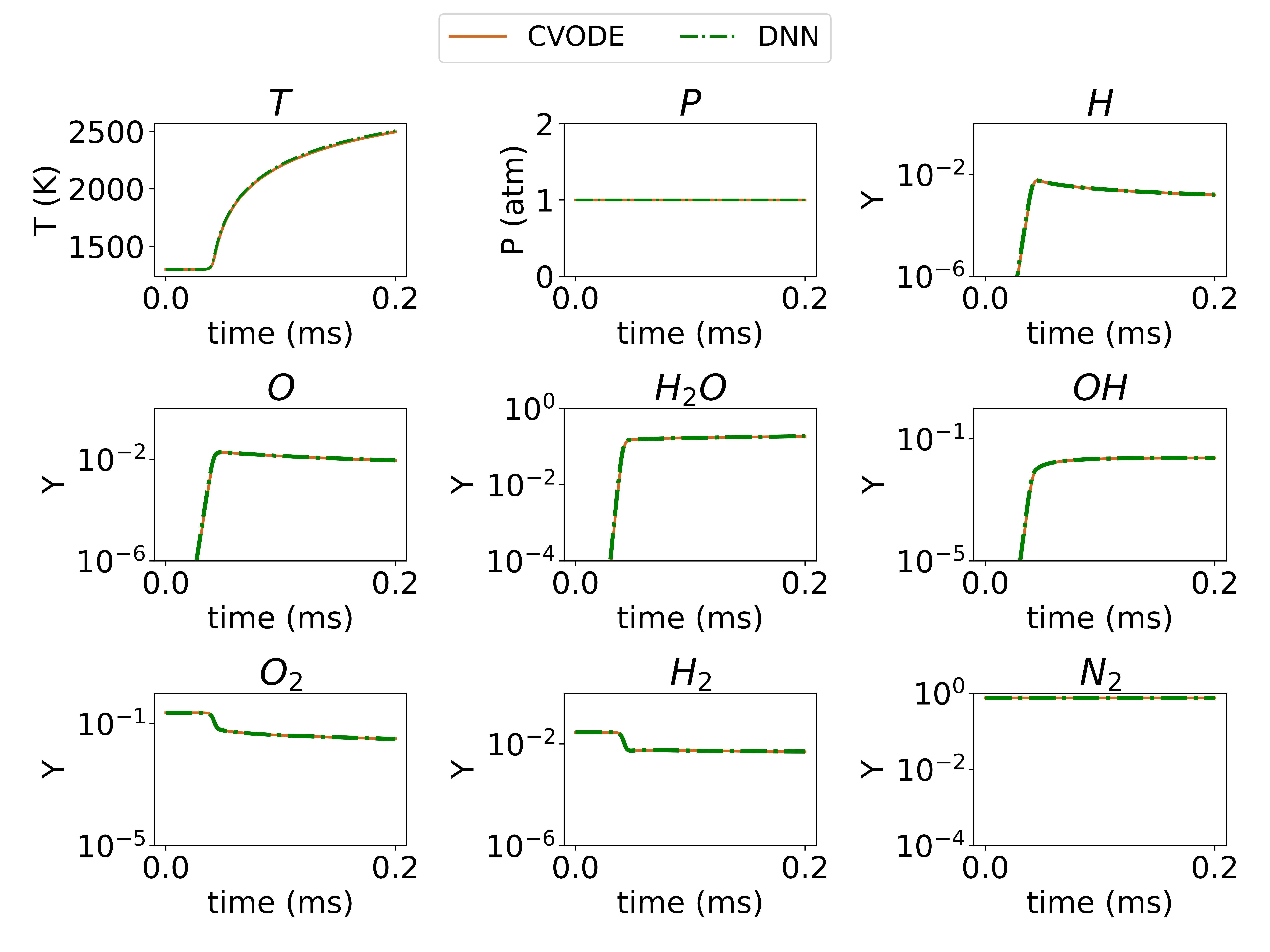}
	}
	\subfigure[]
	{ 
		\includegraphics[width=0.55\textwidth]{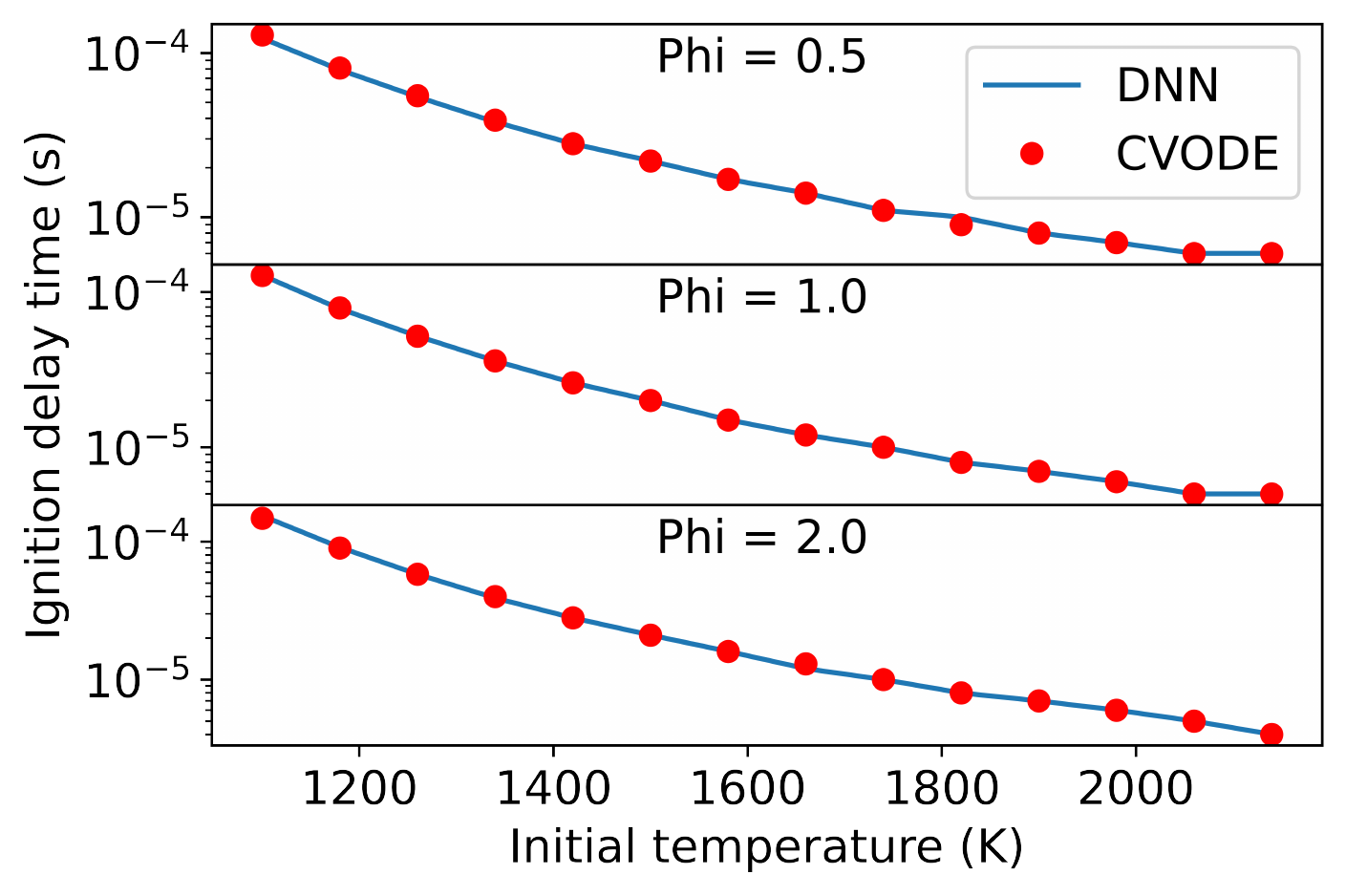}
	}
	\fi
	\caption{Zero-dimensional constant-pressure autoignition results comparison between Cantera and DNN. (a). autoignition at $\phi = 1.0$, T = 1300 K, P = 1 atm; (b). Ignition delay time comparison at P = 1 atm.  
	}
	\label{fig:0d}
\end{figure}

Figure \ref{fig:0d} compares zero-dimensional constant-pressure autoignition results by CVODE and DNN. Figure \ref{fig:0d}a shows the constant-pressure autoignition results by DNN and Cantera at T = 1300 K, P = 1 atm, $\phi=0.9$. The evolution of temperature, pressure, and mass fractions reveals the satisfying accuracy of the DNN prediction. Based on the evolution results, the time interval for the mixture to reach the maximum heat release point is defined as the ignition delay time. Figure \ref{fig:0d}b shows a detailed comparison of ignition delay time predicted by DNN and Cantera at P = 1 atm. The initial temperature ranges from 1100 K to 2500 K, and the equivalence ratio is 0.5 to 2.0. The overall agreement demonstrates the stability and robustness of the DNN in different initial conditions. 

\begin{figure}[!h]
	\centering
	\ifx\mycmd\undefined
	\subfigure[]
	{ 
		\includegraphics[width=0.48\textwidth]{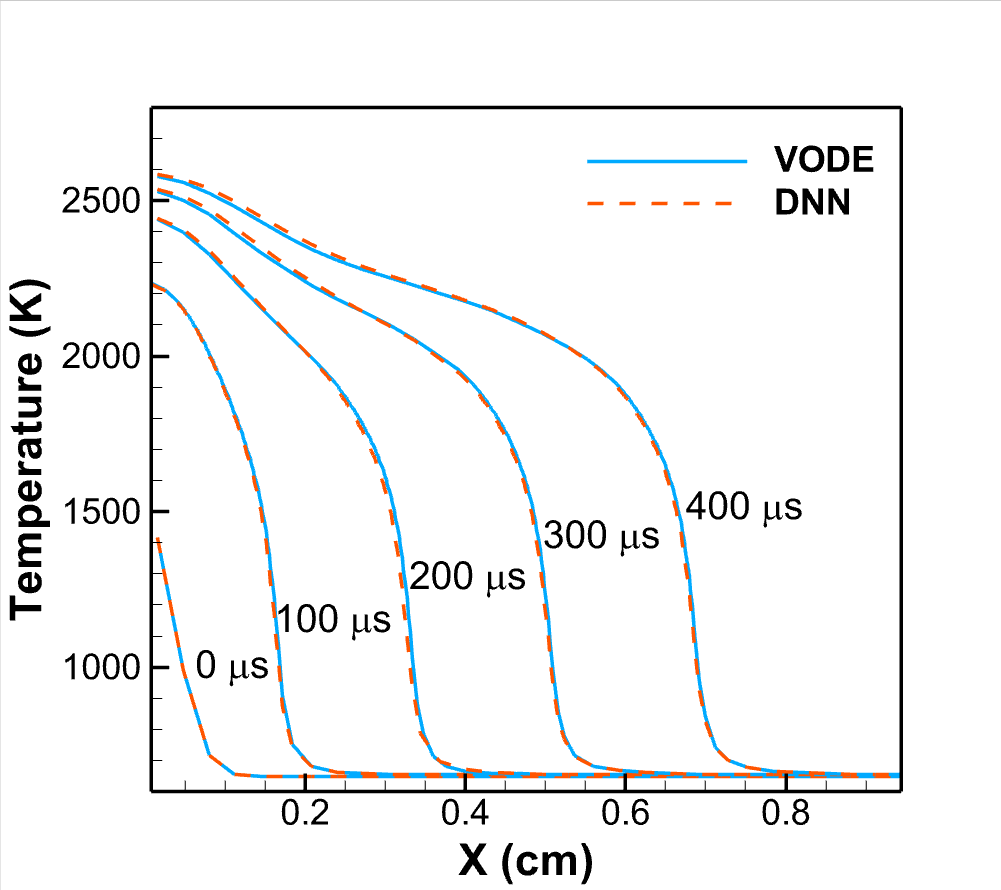}
	}
	\subfigure[]
	{ 
		\includegraphics[width=0.48\textwidth]{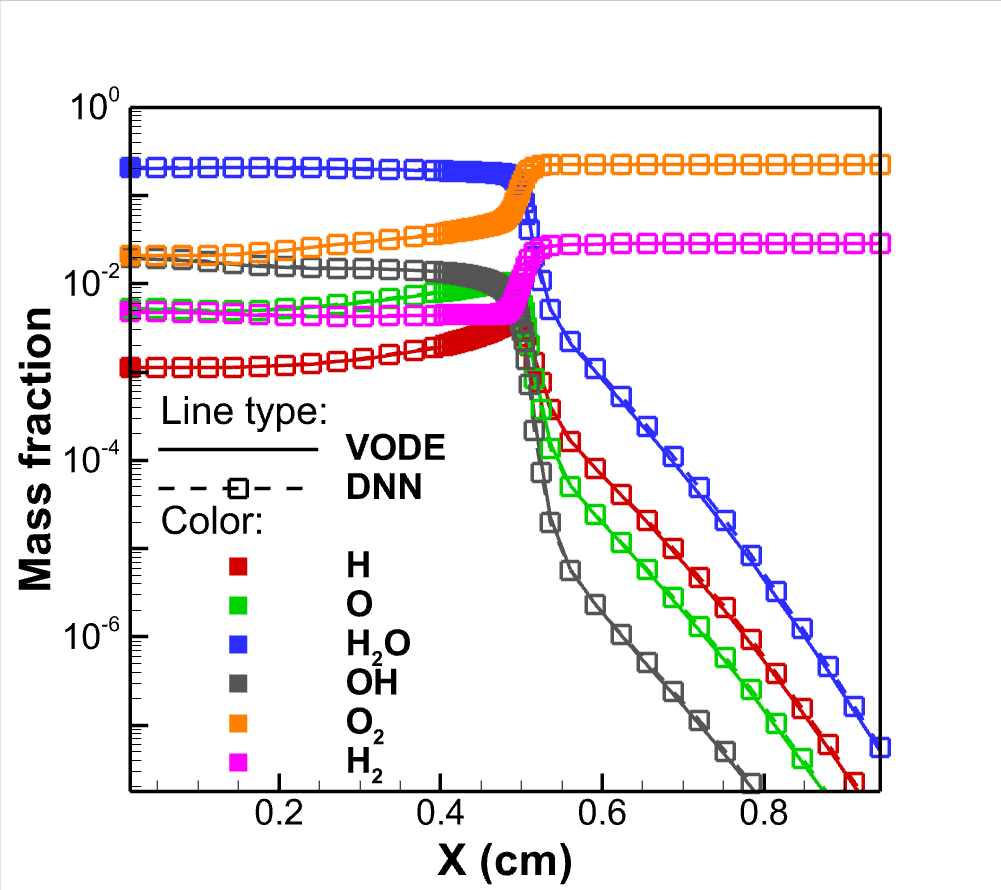}
	}
	\subfigure[]
	{ 
		\includegraphics[width=0.48\textwidth]{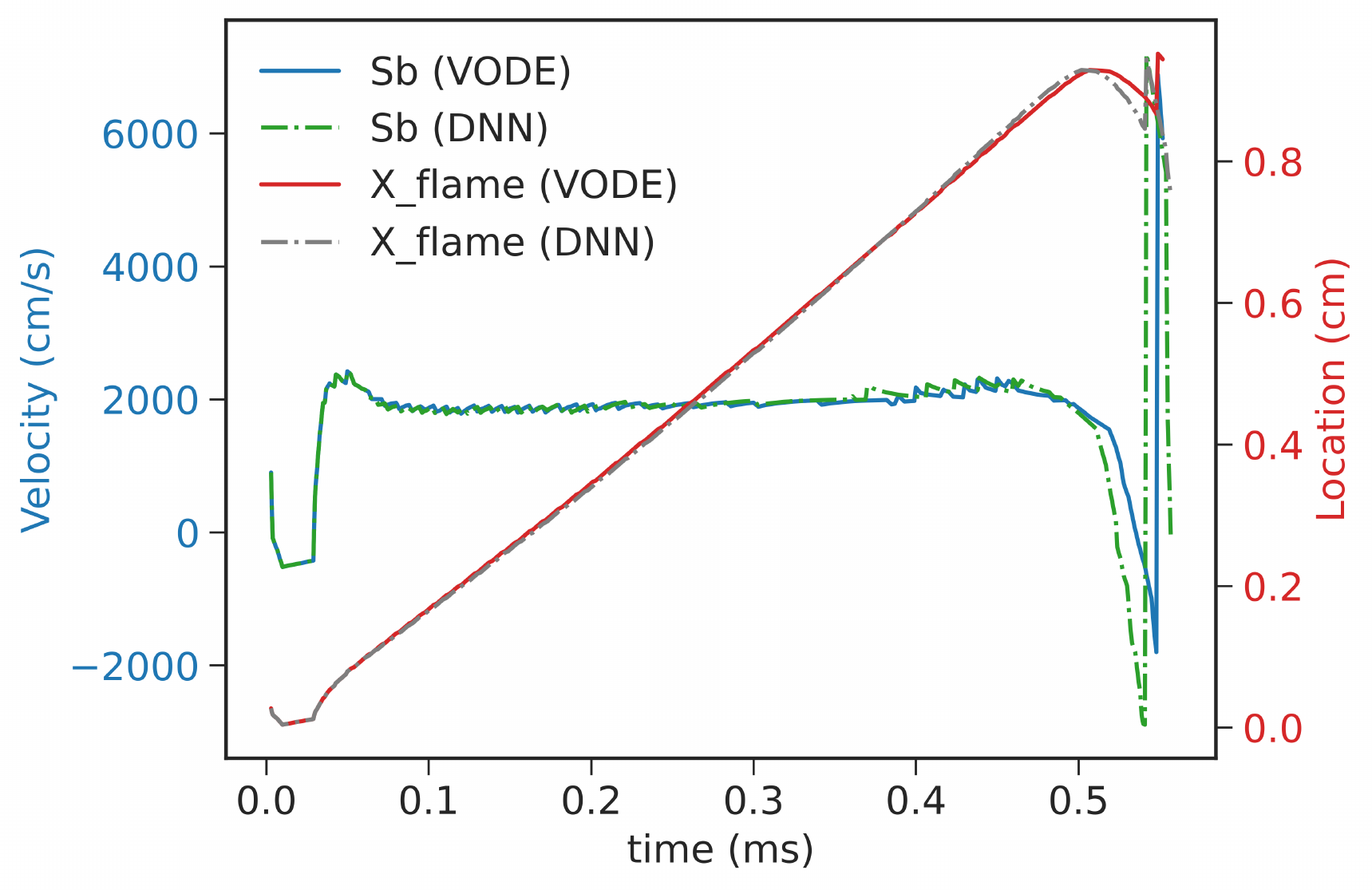}
	}
	\subfigure[]
	{ 
		\includegraphics[width=0.48\textwidth]{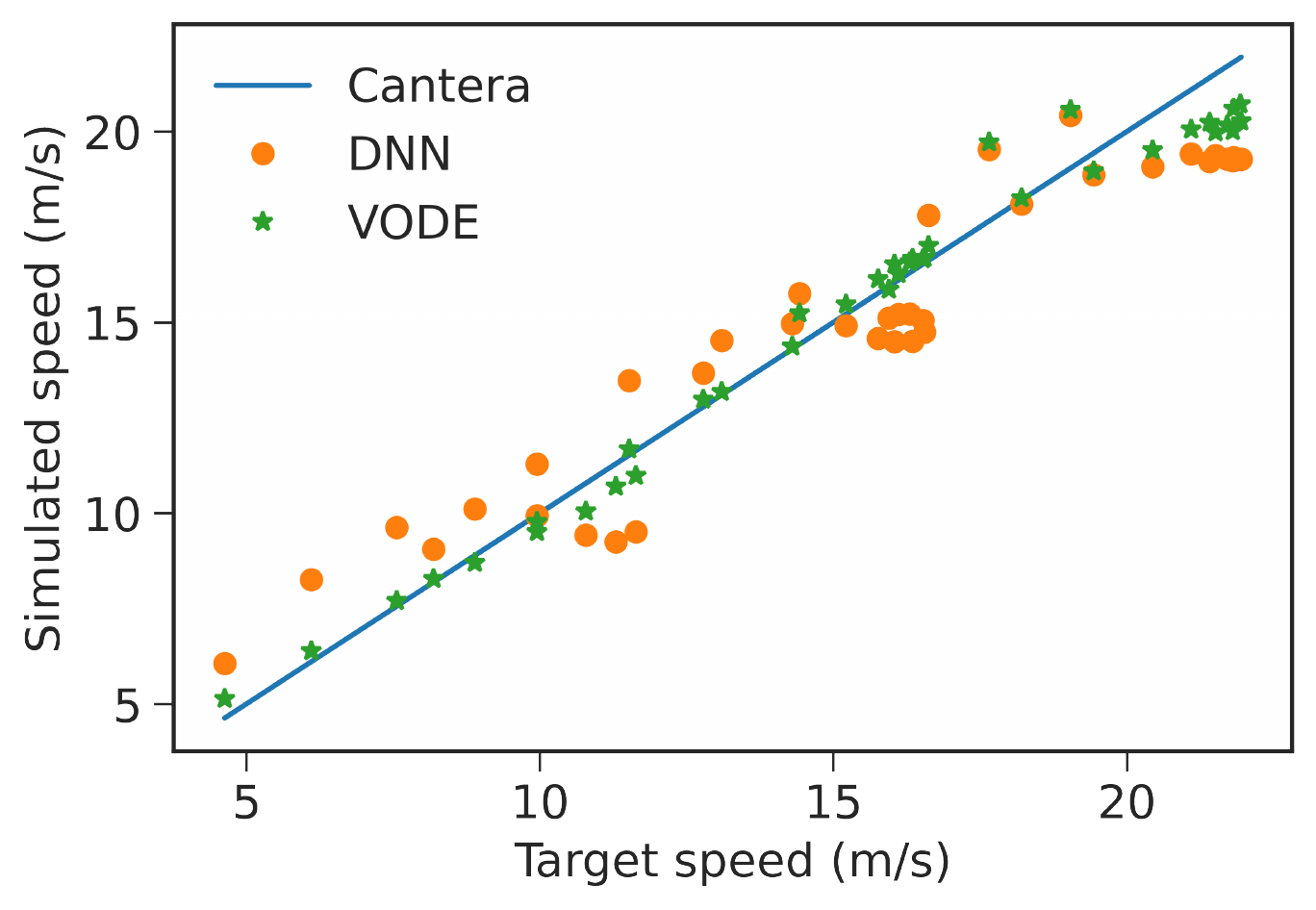}
	}
	\subfigure[]
	{ 
        \includegraphics[width=0.48\textwidth]{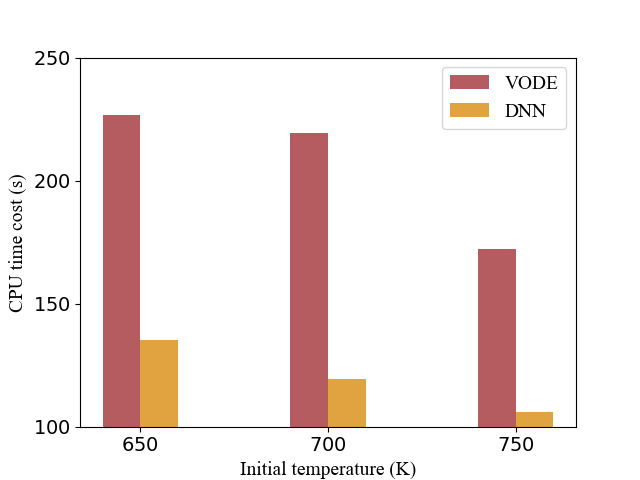}
	}
	\fi
	\caption{One-dimensional transient premixed flame propagation comparison using ASURF with VODE or DNN as the ODE solver. \zth{(a). Flame temperature distribution comparison at $t = 0, 100, 200, 300, 400 \mu s$, $\phi = 1.0$, T = 650 K, P = 1 atm; (b). Species mass fraction comparison at $t = 300  \mu s$, $\phi = 1.0$, T = 650 K, P = 1 atm; (c). Flame trajectory and flame speed $S_b$ at $\phi = 1.1$, T = 650 K, P = 1 atm; (d). Laminar flame speed comparison between ASURF-VODE, ASURF-DNN, and Cantera for initial conditions $T\in [300K, 800K] , \phi\in[0.5,2.0], P = 1 atm$;} \rev{(e). CPU time cost comparison of the ODE integration between ASURF-VODE and ASURF-DNN for initial conditions $T = 650, 700, 750 K, \phi = 1, P = 1 atm$.}  
    }
	\label{fig:1d}
\end{figure}

Figure \ref{fig:1d} compares the transient laminar flame propagation results by ASURF \cite{chen2009a, Zhang2020e} with VODE or DNN as the ODE integrator. Since the training dataset of the DNN does not cover the initial conditions below 800 K, where the stiffness of the chemical system is not a major issue, the DNN will only predict the grid with a temperature higher than 800 K, the rest of the grids using VODE. \zth{For the simulation setup, the initial condition is uniform except for a high-temperature kernel on the left boundary to mimic the spark ignition. Figure \ref{fig:1d}a shows the temporal evolution of temperature distribution for a flame with initial condition $\Phi = 1.0$, T = 650 K, P = 1 atm at t = $0, 100, 200, 300, 400\mu s$. The blue solid line is the VODE result and the orange dotted line is the DNN result. The comparison reveals the satisfying accuracy achieved by DNN in the whole propagation process, including the ignition and stable propagation stages. Figure \ref{fig:1d}b shows a more detailed comparison of species mass fractions to depict the flame structure at t = $300\mu s$ with the initial condition at $\Phi = 1.0$, T = 650 K, P = 1 atm. The level of accuracy agrees with Figure \ref{fig:1d}a that DNN predicts well for all species.} Figure \ref{fig:1d}c shows the temporal evolutions of flame front location and propagation speed at T = 650 K, P = 1 atm, $\phi = 1.1$ using DNN or VODE as the integrator. Two ODE integrators’ results agree well except for a small domain when the flame is near the outlet. Figure \ref{fig:1d}d shows the statistical performance of the DNN prediction coupled with a CFD solver. The DNN prediction  is reasonably accurate: averaged laminar flame speed error is 6.9\%, and the maximum error is around 20\%. There are several possible reasons behind the error. The main reason might be the DNN prediction error accumulation. Since the DNN needs to provide predictions continuously, some small errors might transport to the neighboring grid and accumulate during the temporal evolution. The accumulation of prediction errors can lead to sensible errors in laminar flame speed. \rev{Figure \ref{fig:1d}e compares the CPU time cost on the ODE integration between the VODE method and the DNN. The efficiency improvement is above 70\%. Considering the relatively small size of the hydrogen/air mechanism, the current acceleration indicates a promising future of the DNN method for more complex fuels with stronger chemical stiffness.}
\subsection{Multi-dimensional lifted jet flame simulations}
In order to demonstrate the capabilities of the DNN model under complex flow conditions, 2D laminar and 3D turbulent non-premixed lifted jet flames are simulated. It is worth recalling that the DNN was trained \textit{a priori} without knowing any specific flow information. Hence, the robustness of the proposed multi-scale sampling method can be properly assessed using these higher-dimensional test cases. A finite volume in-house code with third-order accuracy for both space and time is used to solve the Navier-Stokes equations with detailed chemical kinetics and mixture-averaged transport (see detailed description and validation in Supplementary Material). 

\begin{figure}[!h]
	\centering
	\ifx\mycmd\undefined
	\includegraphics[width=0.6\textwidth]{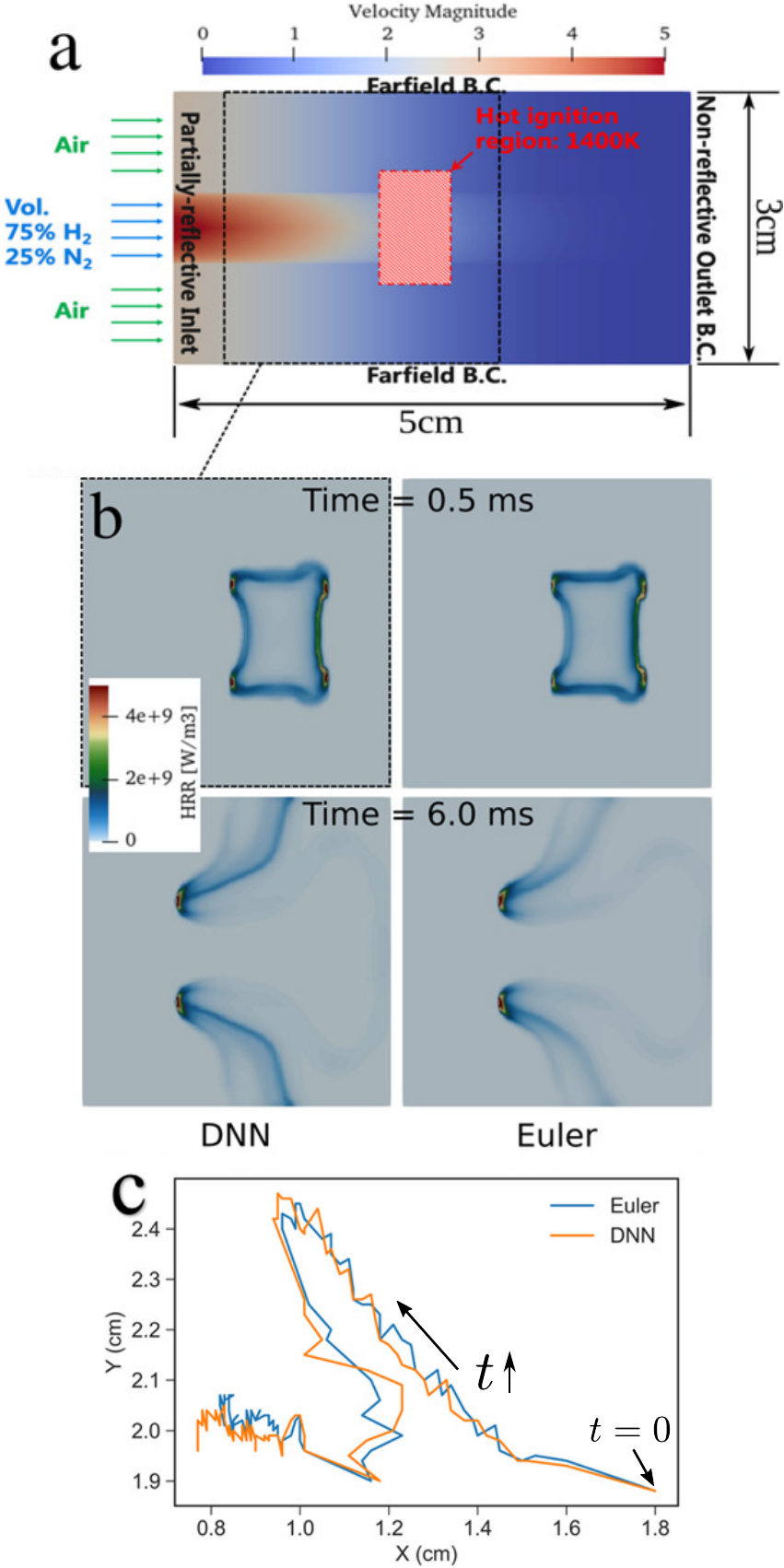}
	\fi
	\caption{Two-dimensional jet flame simulation results using DNN and implicit Euler solver. (a) computational setup; (b) heat release rate contours at $t = 0.5$ and 6.0 ms; (c) temporal evolution of the triple point.
	}
	\label{fig:2djet}
\end{figure}

Figure \ref{fig:2djet} compares the 2D laminar planar jet flame simulation results obtained using DNN and implicit Euler as integrators. The computational setup is schematically shown in Figure \ref{fig:2djet}a. It consists of a $3\times 5$ cm domain with a H$_2$ jet surrounded by an air co-flow at an initial pressure of 1 atm and temperature of 300 K. The jet diameter is 8 mm. The inlet mixture is 75\% H$_2$ and 25\% N$_2$ by volume with a velocity of 5 m/s and the air coflow velocity is 3 m/s. An initial high-temperature region at T = 1400 K is set at the center of the domain to initiate the flame. Figure \ref{fig:2djet}b shows the evolution of the heat release rate (HRR) contours for two instants. It can be seen that two flame branches form on each side of the jet soon after the ignition at 0.5 ms, and the reaction zone is almost identical for the DNN and Euler cases. Later at 6 ms, the well-known \textit{triple flame} structure \cite{Ruetsch1995} is formed and the two cases still show overall very good agreement despite some slight difference in the HRR magnitude along the tail of the middle diffusion branch. Furthermore, the trajectories of the triple point from ignition to stabilization given by the DNN and Euler solvers are presented in Figure \ref{fig:2djet}c showing a reasonably good agreement.
%Since the current DNN does not cover the low-temperature domain below 800 K, the corresponding domain is calculated using implicit Euler. 
This suggests that the DNN is able to capture the transient evolution of the triple flame: the distinct characteristics of the diffusion flame, lean and rich premixed flames are intrinsically predicted by the trained model.  
\begin{figure}[!h]
	\centering
	\ifx\mycmd\undefined
	\includegraphics[width=0.6\textwidth]{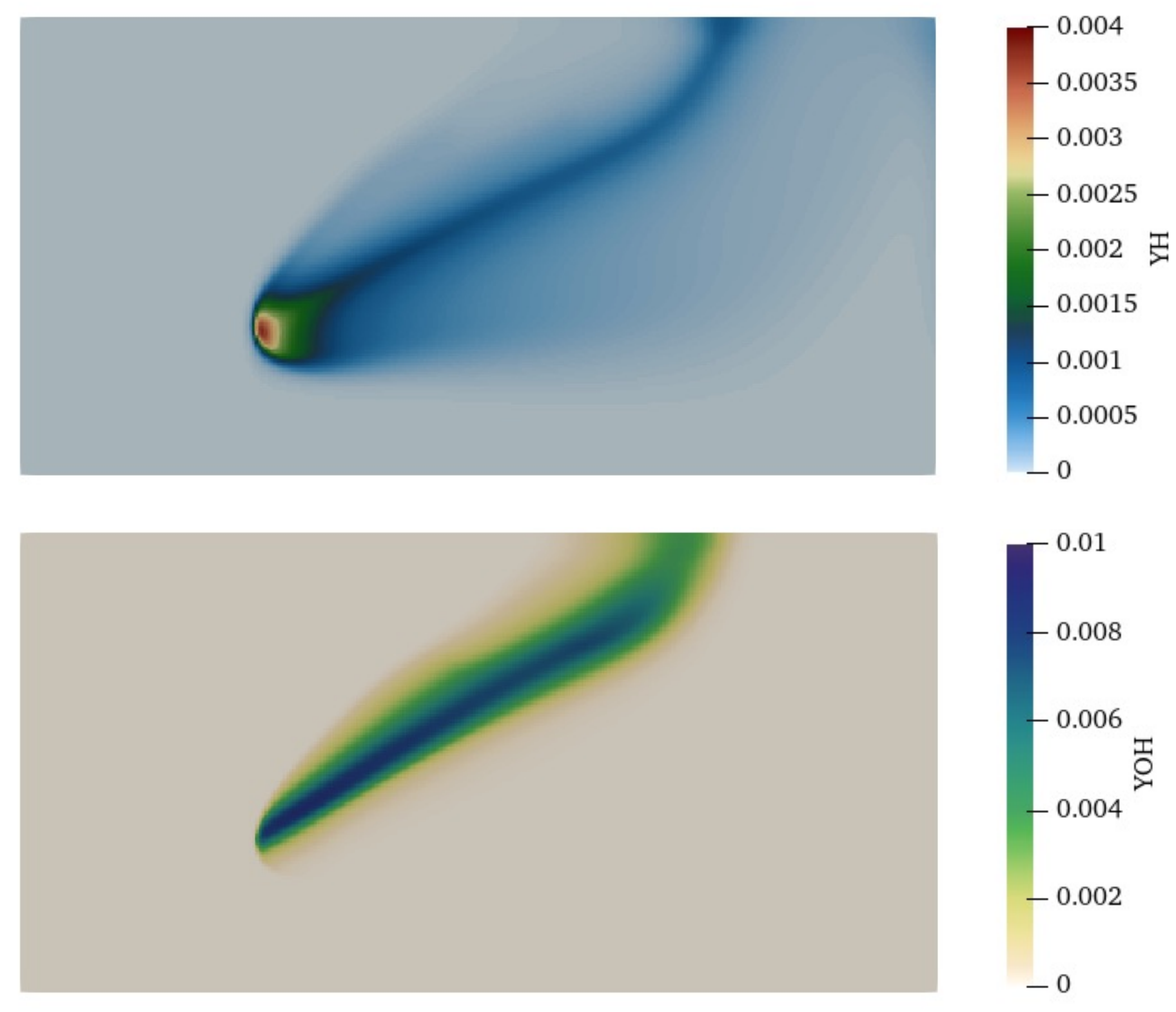}
	\fi
	\caption{DNN prediction for H and OH radials of 2D laminar triple flame.
	}
	\label{fig:2djet-H-OH}
\end{figure}

\begin{figure}[!h]
	\centering
% 	\ifx\mycmd\undefined
	\includegraphics[width=0.7\textwidth]{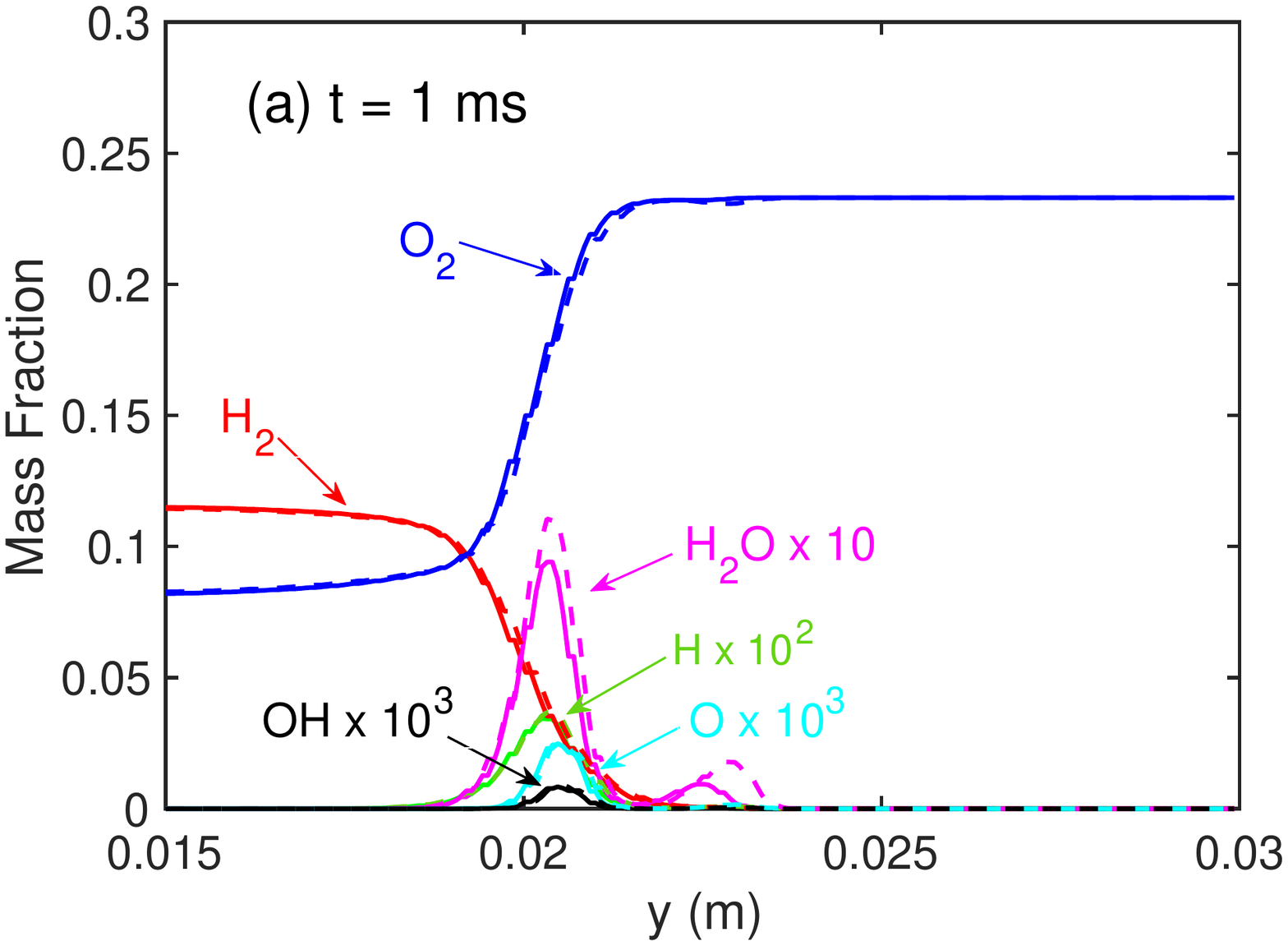}
	\includegraphics[width=0.7\textwidth]{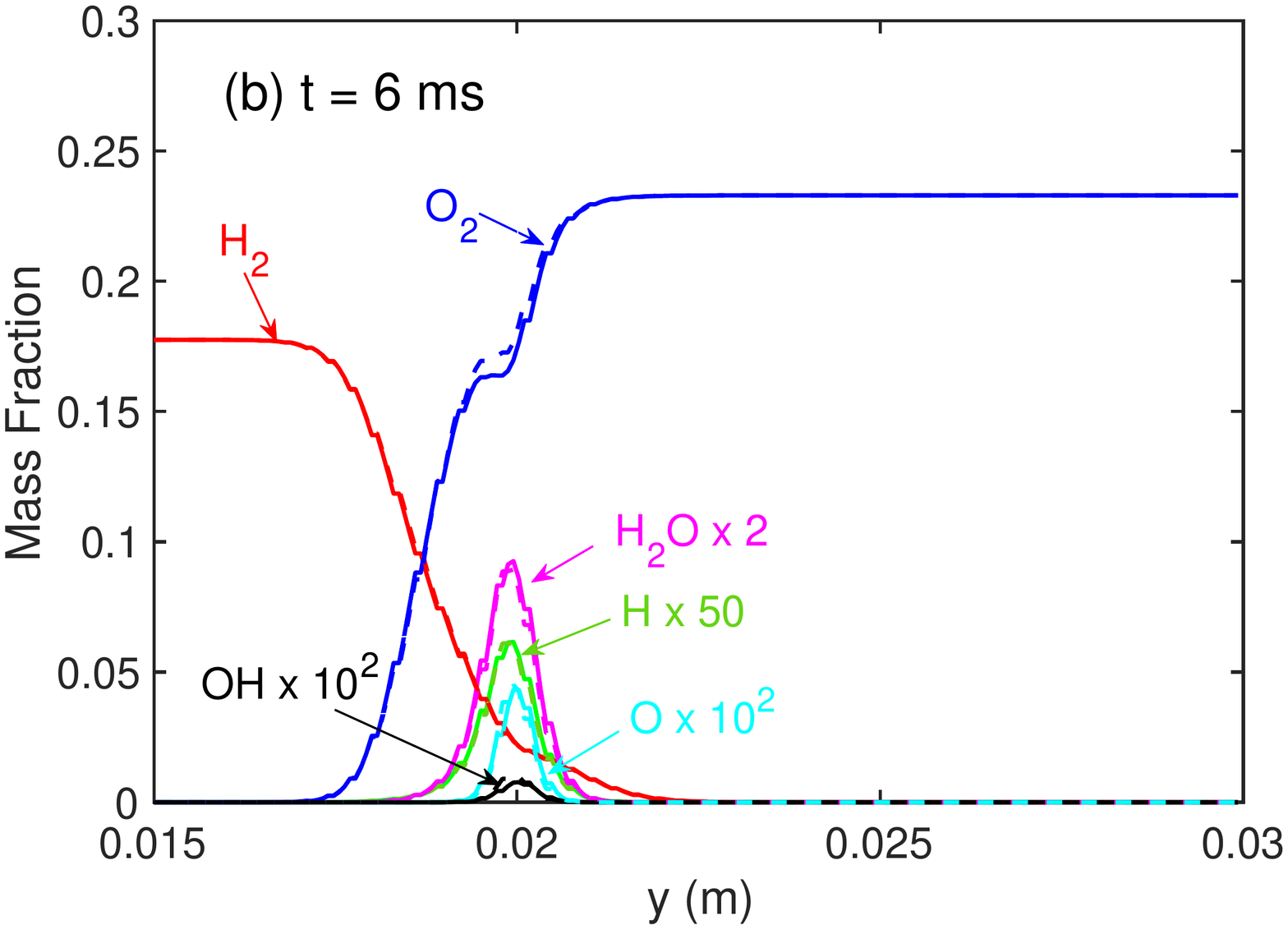}
% 	\fi
	\caption{\zxc{Radial profiles of species mass fractions at the maximum heat release point 2D laminar triple flame obtained using DNN (dashed) and Euler (solid) solvers.}
	}
	\label{fig:Yprofile}
\end{figure}

Figure \ref{fig:2djet-H-OH} depicts the distributions of H and OH radical mass fractions at the stabilization height for $t = 10$ ms for the DNN case. It is shown that all three reaction branches of the triple flame are well captured using the H atom, whereas the diffusion branch is distinctly represented by the OH radicals. 
\zxc{In order to make a quantitative comparison, radial profiles of species mass fractions are plotted in Figure~\ref{fig:Yprofile} for two time instants at $t = 1$ and $6$~ms. The axial position of the maximum heat release rate point is chosen for these plots. At both times, the results given by the Euler and DNN solvers show very good agreement for all major and minor species.} These results further confirm that the fine interaction between the detailed chemical kinetics and the jet flow is correctly handled by the DNN model.  

So far the DNN model has shown good performance for predicting chemical kinetics near the main reaction manifold with little or moderate flow field perturbations in homogeneous mixtures, 1D and 2D laminar flames. 
Next, we further test the DNN in a 3D highly turbulent case with strong variations in the local thermochemical states. The experiment of Cheng et al. \cite{cheng1992} for hydrogen lifted flame stabilized in quiescent air at room temperature and pressure is simulated.  
The round jet diameter was $D = 2$ mm and pure H$_2$ fuel was injected at 680 m/s with a corresponding Mach number of 0.58 and Reynolds number of 13,600. 
The computational setup (mesh resolution, domain size, etc.) from a previous DNS \cite{Mizobuchi2002a} is adopted here and the grid size is $800 \times 500 \times 500$ in the streamwise and cross-stream directions.
It is worth remarking that the time-step size of 20 ns used in the simulation with DNN is an order of magnitude larger than that required for the implicit ODE solver since the chemistry stiffness is already treated in the DNN. Note that the unit cost for a single-step DNN prediction is about 2 times higher. This is because each prediction essentially performs a multiplication between two large dense matrices. Overall, there is still a considerable reduction in the computational cost by a factor of 2 to 3 by using the DNN. The further speedup can be achieved by involving dynamic load balancing and potentially GPU architectures \cite{Hernandez2018, Lu2021}, as accelerating matrix operation is straightforward compared to logical iterations in conventional approaches. This will be explored in our future works.   
\begin{figure}[th!]
    \centering
	\ifx\mycmd\undefined
	\includegraphics[width=0.5\textwidth]{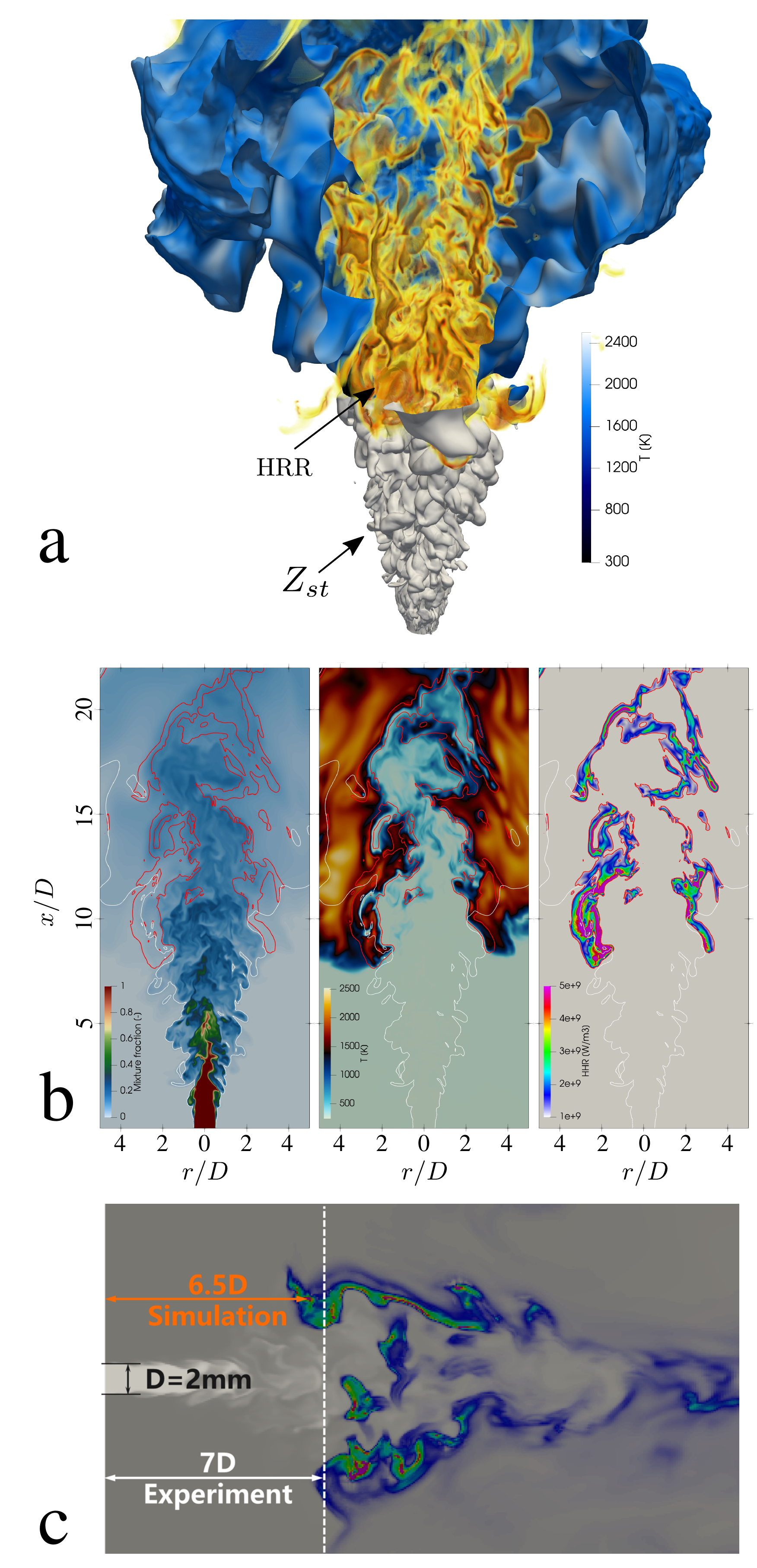}
	\fi
	\caption{Three-dimensional turbulent lifted flame simulation using DNN chemistry solver. (a) volume-rendered HRR (yellow) and stoichiometric mixture fraction iso-surface colored by temperature (blue); (b) instantaneous $x$-$r$ slice contours for mixture fraction (left), temperature (middle) and HRR (right); (c) comparison of the final lift-off height between the simulation and experiment. 
	}
	\label{fig:3djet}
\end{figure}

The simulation was first run on a coarse grid for the initial flame kernel development and the solution was then mapped to the refined grid for the flame stabilization period. The ignition was initiated by placing a high-temperature sphere with a radius of 2 mm at 7D downstream of the jet exit. 
Figure \ref{fig:3djet}a shows 3D rendering of the evolving flame on the refined mesh at about 1.5 ms after ignition. 
It is seen that while most of the heat release occurs near the stoichiometric iso-surface, the finger-like structures indicating the lean premixed branches are visible in the low-velocity regions around the jet. These are unique for lifted flames stabilized in quiescent air without any hot coflows, also consistent with the findings from the previous DNS study \cite{Mizobuchi2002a}. The 2D mid-plane contours of mixture fraction, temperature, and HRR are presented in Figure \ref{fig:3djet}b, exhibiting generally consistent flame behaviors one would expect from such lifted flame configurations. 
Finally, the flame lift-off height, a quantity measuring the overall complex interaction between the propagating triple flame and turbulent incoming flow, is compared between the simulation and experiment in Figure \ref{fig:3djet}c to assess the DNN model accuracy quantitatively. The averaged height measured was about 7D \cite{cheng1992} and the simulated value fluctuates between 6 and 7D, indicating a good agreement with the experiment. 
\zxc{Detailed comparison of the flame statistics is required to further validate the proposed method, for which a full DNS run time needs to be performed covering a large number of data snapshots. This is beyond the scope of this work and will be conducted in a future study.
Nonetheless, this 3D turbulent lifted flame test case demonstrates the capabilities of the offline-trained DNN model to predict chemical kinetics in DNS of turbulent combustion, showing very good prediction accuracy with a considerable reduction in computational cost.}
It also suggests that a simple DNN structure with an appropriate sampling strategy has great potential for accommodating detailed combustion chemistry in real-world turbulent flame simulations at moderate costs. 

\section{Conclusions} 
The major advantage of the DNN-based model is no time step size limit for the DNN to predict the chemical kinetics. It can be a powerful tool to overcome stiffness problems in combustion ODE integration. The key question is how to train a DNN to be accurate and generic.

The present work explores a preprocessing and multi-scale sampling method to improve DNN prediction performance for combustion chemical kinetics. The golden rule for the neural network is always that data selection determines neural network performance. The results verify that the deep neural networks trained on different datasets have divergent prediction abilities, even though they have the same neural network structure. 
More specifically, the DNN trained on the manifold dataset obtained from a given combustion configuration (i.e., PSR, 1D laminar flame, etc.) can accurately predict the evolution of data points, that is, thermochemical states on the same manifold. However, the sampling and prediction are limited to the specific manifold and can hardly be applied in a generalized scenario for practical combustor simulation. The Monte Carlo-based dataset is independent of combustion configuration and consists of random data in a wide thermochemical phase space. Yet, it cannot predict small-scale data well. The large relative prediction error prevents the corresponding DNN from being applicable in any continuous evolution cases.

The multi-scale sampling and the Box-Cox transformation proposed in this work can overcome the difficulty of collecting multi-scale combustion data, leading to a DNN model with accurate and robust predictions. The data generation does not rely on the type of combustion configuration. A dataset with 5,300,000 samples is generated covering a large phase space: $T\in\left[800K,3100K\right], P\in\left[0.5atm,\ 2atm\right]$, and no limitations on the equivalence ratio. Subsequently, a systematic and comprehensive validation is performed using a broad range of test cases, including autoignition of 0D homogeneous mixtures, 1D laminar flame propagation, 2D laminar jet flame with triple-flame structure, and 3D turbulent lifted flame. The quantitative comparison confirms the deep neural network's robustness and reasonable accuracy for predicting the ignition delay time, laminar flame speed, triple point trajectory, and the flame lift-off height. It is remarkable that the same DNN model, without any specific tuning, can correctly represent the chemical kinetics and its interaction with various flow features across such a wide range of configurations. 

The DNN model and the example Fortran and Python codes are provided to cross-validate the conclusions. Any improvement in the future can be directly compared through routine benchmark cases such as ignition delay time and laminar flame speed. Future works are planned towards more complex hydrocarbon fuels and integrated learning for a combined chemical mechanism reduction and kinetics prediction. 

\vspace{1cm}
\section*{Acknowledgements} \addvspace{10pt}
This work is supported by the Shanghai Sailing Program, the Natural Science Foundation of Shanghai Grant No. 20ZR1429000  (Z. X.), the National Natural Science Foundation of China Grant No. 62002221 (Z. X.), the National Natural Science Foundation of China Grant No. 12101402 (Y. Z.), Shanghai Municipal of Science and Technology Project Grant No. 20JC1419500 (Y.Z.), Shanghai Municipal of Science and Technology Major Project No. 2021SHZDZX0102 (Z. X., Y.Z.), and the HPC of School of Mathematical Sciences (Z. X., Y.Z.) and the Student Innovation Center (Z. X., Y.Z.) at Shanghai Jiao Tong University, and AI for Science Institute, Beijing.
%\noindent \acknowledge{Acknowledgments} 

% -------------------------------------------------------------------- %
% -------------------------------------------------------------------- %
% -------------------------------------------------------------------- %

% -------------------------------------------------------------------- %
% -------------------------------------------------------------------- %
% -------------------------------------------------------------------- %
\bibliographystyle{pci}
\noindent \bibliography{library}

\begin{thebibliography}{10}
\expandafter\ifx\csname url\endcsname\relax
  \def\url#1{\texttt{#1}}\fi
\expandafter\ifx\csname urlprefix\endcsname\relax\def\urlprefix{URL }\fi
\expandafter\ifx\csname href\endcsname\relax
  \def\href#1#2{#2} \def\path#1{#1}\fi

\bibitem{Lu2009b}
T.~Lu, C.~K. Law,
  \href{https://ac.els-cdn.com/S036012850800066X/1-s2.0-S036012850800066X-main.pdf?_tid=ffd6bd25-7f3b-431e-93e2-f45f51ffe4dd&acdnat=1530807627_d3ec4500ab3df5b9621c5bf061c19862
  https://linkinghub.elsevier.com/retrieve/pii/S036012850800066X}{{Toward
  accommodating realistic fuel chemistry in large-scale computations}},
  Progress in Energy and Combustion Science 35~(2) (2009) 192--215.
\newblock

\bibitem{Christo1996}
F.~Christo, A.~R. Masri, E.~M. Nebot,
  \href{https://linkinghub.elsevier.com/retrieve/pii/0010218095002502}{{Artificial
  neural network implementation of chemistry with pdf simulation of H2/CO2
  flames}}, Combustion and Flame 106~(4) (1996) 406--427.
\newblock

\bibitem{Christo1996a}
F.~Christo, A.~Masri, E.~Nebot, S.~Pope,
  \href{https://linkinghub.elsevier.com/retrieve/pii/S0082078496801986}{{An
  integrated PDF/neural network approach for simulating turbulent reacting
  systems}}, Symposium (International) on Combustion 26~(1) (1996) 43--48.
\newblock

\bibitem{Blasco1998}
J.~A. Blasco, N.~Fueyo, C.~Dopazo, J.~Ballester, {Modelling the temporal
  evolution of a reduced combustion chemical system with an artificial neural
  network}, Combustion and Flame 113~(1-2) (1998) 38--52.
\newblock

\bibitem{Blasco2000}
J.~Blasco, N.~Fueyo, C.~Dopazo, J.-Y. Chen,
  \href{http://www.tandfonline.com/doi/abs/10.1088/1364-7830/4/1/304}{{A
  self-organizing-map approach to chemistry representation in combustion
  applications}}, Combustion Theory and Modelling 4~(1) (2000) 61--76.
\newblock

\bibitem{Chen2000}
J.-Y. Chen, J.~Blasco, N.~Fueyo, C.~Dopazo,
  \href{https://linkinghub.elsevier.com/retrieve/pii/S0082078400802027}{{An
  economical strategy for storage of chemical kinetics: Fitting in situ
  adaptive tabulation with artificial neural networks}}, Proceedings of the
  Combustion Institute 28~(1) (2000) 115--121.
\newblock

\bibitem{Kempf2005}
A.~Kempf, F.~Flemming, J.~Janicka,
  \href{http://dx.doi.org/10.1016/j.proci.2004.08.182}{{Investigation of
  lengthscales , scalar dissipation , and flame orientation in a piloted
  diffusion flame by LES}}, Proceedings of the Combustion Institute 30~(1)
  (2005) 557--565.
\newblock

\bibitem{Ihme2009}
M.~Ihme, C.~Schmitt, H.~Pitsch,
  \href{http://dx.doi.org/10.1016/j.proci.2008.06.100
  https://linkinghub.elsevier.com/retrieve/pii/S1540748908001132}{{Optimal
  artificial neural networks and tabulation methods for chemistry
  representation in LES of a bluff-body swirl-stabilized flame}}, Proceedings
  of the Combustion Institute 32~(1) (2009) 1527--1535.
\newblock

\bibitem{Sen2009}
B.~A. Sen, S.~Menon,
  \href{http://dx.doi.org/10.1016/j.proci.2008.05.077}{{Turbulent premixed
  flame modeling using artificial neural networks based chemical kinetics}},
  Proceedings of the Combustion Institute 32 I~(1) (2009) 1605--1611.
\newblock

\bibitem{Sen2010}
B.~A. Sen, S.~Menon,
  \href{http://dx.doi.org/10.1016/j.combustflame.2009.06.005}{{Linear eddy
  mixing based tabulation and artificial neural networks for large eddy
  simulations of turbulent flames}}, Combustion and Flame 157~(1) (2010)
  62--74.
\newblock

\bibitem{Chatzopoulos2013}
A.~Chatzopoulos, S.~Rigopoulos,
  \href{http://dx.doi.org/10.1016/j.proci.2012.06.057
  https://linkinghub.elsevier.com/retrieve/pii/S1540748912001654}{{A chemistry
  tabulation approach via Rate-Controlled Constrained Equilibrium (RCCE) and
  Artificial Neural Networks (ANNs), with application to turbulent non-premixed
  CH4/H2/N2 flames}}, Proceedings of the Combustion Institute 34~(1) (2013)
  1465--1473.
\newblock

\bibitem{Franke2017}
L.~L. Franke, A.~K. Chatzopoulos, S.~Rigopoulos,
  \href{http://dx.doi.org/10.1016/j.combustflame.2017.07.014
  https://linkinghub.elsevier.com/retrieve/pii/S0010218017302596}{{Tabulation
  of combustion chemistry via Artificial Neural Networks (ANNs): Methodology
  and application to LES-PDF simulation of Sydney flame L}}, Combustion and
  Flame 185 (2017) 245--260.
\newblock

\bibitem{Wan2020}
K.~Wan, C.~Barnaud, L.~Vervisch, P.~Domingo,
  \href{https://linkinghub.elsevier.com/retrieve/pii/S0010218020302170
  https://doi.org/10.1016/j.combustflame.2020.06.008}{{Chemistry reduction
  using machine learning trained from non-premixed micro-mixing modeling:
  Application to DNS of a syngas turbulent oxy-flame with side-wall effects}},
  Combustion and Flame 220 (2020) 119--129.
\newblock

\bibitem{Ding2021}
T.~Ding, T.~Readshaw, S.~Rigopoulos, W.~P. Jones,
  \href{https://doi.org/10.1016/j.combustflame.2021.111493}{{Machine learning
  tabulation of thermochemistry in turbulent combustion: An approach based on
  hybrid flamelet/random data and multiple multilayer perceptrons}}, Combustion
  and Flame 231 (2021) 111493.
\newblock

\bibitem{Chi2021}
C.~Chi, G.~Janiga, D.~Th{\'{e}}venin,
  \href{https://linkinghub.elsevier.com/retrieve/pii/S0010218020305939}{{On-the-fly
  artificial neural network for chemical kinetics in direct numerical
  simulations of premixed combustion}}, Combustion and Flame 226 (2021)
  467--477.
\newblock

\bibitem{Nakazawa2022}
R.~Nakazawa, Y.~Minamoto, N.~Inoue, M.~Tanahashi,
  \href{https://doi.org/10.1016/j.combustflame.2021.111696
  https://linkinghub.elsevier.com/retrieve/pii/S0010218021004399}{{Species
  reaction rate modelling based on physics-guided machine learning}},
  Combustion and Flame 235~(xxxx) (2022) 111696.
\newblock

\bibitem{Sutherland2009}
J.~C. Sutherland, A.~Parente,
  \href{http://dx.doi.org/10.1016/j.proci.2008.06.147}{{Combustion modeling
  using principal component analysis}}, Proceedings of the Combustion Institute
  32 I~(1) (2009) 1563--1570.
\newblock

\bibitem{Malik2021}
M.~R. Malik, P.~{Obando Vega}, A.~Coussement, A.~Parente, {Combustion modeling
  using Principal Component Analysis: A posteriori validation on Sandia flames
  D, E and F}, Proceedings of the Combustion Institute 38~(2) (2021)
  2635--2643.
\newblock

\bibitem{Mirgolbabaei2014}
H.~Mirgolbabaei, T.~Echekki,
  \href{http://dx.doi.org/10.1016/j.combustflame.2013.08.016}{{Nonlinear
  reduction of combustion composition space with kernel principal component
  analysis}}, Combustion and Flame 161~(1) (2014) 118.
\newblock

\bibitem{DAlessio2020}
G.~D'Alessio, A.~Parente, A.~Stagni, A.~Cuoci, G.~D. Alessio, A.~Parente,
  A.~Stagni, A.~Cuoci,
  \href{https://doi.org/10.1016/j.combustflame.2019.09.010}{{Adaptive chemistry
  via pre-partitioning of composition space and mechanism reduction}},
  Combustion and Flame 211 (2020) 68--82.
\newblock

\bibitem{zhu2017unpaired}
J.-Y. Zhu, T.~Park, P.~Isola, A.~A. Efros, Unpaired image-to-image translation
  using cycle-consistent adversarial networks, in: Proceedings of the IEEE
  international conference on computer vision, 2017, pp. 2223--2232.

\bibitem{Ju2015a}
Y.~Ju, W.~Sun,
  \href{https://ac.els-cdn.com/S0360128514000781/1-s2.0-S0360128514000781-main.pdf?_tid=be0ed1aa-b04a-11e7-a37c-00000aab0f27&acdnat=1507922252_c18464becf2c6d78f7fd238b5f8d1634
  https://linkinghub.elsevier.com/retrieve/pii/S0360128514000781}{{Plasma
  assisted combustion: Dynamics and chemistry}}, Progress in Energy and
  Combustion Science 48 (2015) 21--83.
\newblock

\bibitem{Zhang2021}
T.~Zhang, A.~J. Susa, R.~K. Hanson, Y.~Ju, {Two-dimensional simulation of cool
  and double flame formation induced by the laser ignition under shock-tube
  conditions}, in: 12th US National Combustion Meeting, no. May, 2021.

\bibitem{xu_training_2018}
Z.-Q.~J. Xu, Y.~Zhang, Y.~Xiao, Training behavior of deep neural network in
  frequency domain, International Conference on Neural Information Processing
  (2019) 264--274.

\bibitem{xu2019frequency}
Z.-Q.~J. Xu, Y.~Zhang, T.~Luo, Y.~Xiao, Z.~Ma, Frequency principle: Fourier
  analysis sheds light on deep neural networks, Communications in Computational
  Physics 28~(5) (2020) 1746--1767.

\bibitem{Box1964}
G.~E.~P. Box, D.~R. Cox,
  \href{https://onlinelibrary.wiley.com/doi/10.1111/j.2517-6161.1964.tb00553.x}{{An
  Analysis of Transformations}}, Journal of the Royal Statistical Society:
  Series B (Methodological) 26~(2) (1964) 211--243.
\newblock

\bibitem{Hendrycks2016}
D.~Hendrycks, K.~Gimpel, \href{http://arxiv.org/abs/1606.08415}{{Gaussian Error
  Linear Units (GELUs)}} (2016) 1--9\href {http://arxiv.org/abs/1606.08415}
  {\path{arXiv:1606.08415}}.

\bibitem{Maas1992}
U.~Maas, S.~B. Pope,
  \href{http://ac.els-cdn.com/001021809290034M/1-s2.0-001021809290034M-main.pdf?_tid=7951b03a-9332-11e7-b0de-00000aacb362&acdnat=1504723245_0004bb1bcd2dc59860d57ab104a6b7e1
  https://ac.els-cdn.com/001021809290034M/1-s2.0-001021809290034M-main.pdf?_tid=60bdc1b0-8b8}{{Simplifying
  chemical kinetics: Intrinsic low-dimensional manifolds in composition
  space}}, Combustion and Flame 88~(3-4) (1992) 239--264.
\newblock

\bibitem{Zhanga}
T.~Zhang, Y.~Zhang, W.~E, Y.~Ju, {DLODE: a deep learning-based ODE solver for
  chemistry kinetics}, in: AIAA Scitech 2021 Forum, American Institute of
  Aeronautics and Astronautics, Reston, Virginia, 2021.
\newblock \href {http://arxiv.org/abs/2012.12654} {\path{arXiv:2012.12654}},

\bibitem{Evans1980}
J.~S. Evans, C.~J. Schexnayder, {Influence of Chemical Kinetics and Unmixedness
  on Burning in Supersonic Hydrogen Flames}, AIAA Journal 18~(2) (1980)
  188--193.
\newblock

\bibitem{cheng1992}
T.~Cheng, J.~Wehrmeyer, R.~Pitz,
  \href{https://linkinghub.elsevier.com/retrieve/pii/001021809290061S}{{Simultaneous
  temperature and multispecies measurement in a lifted hydrogen diffusion
  flame}}, Combustion and Flame 91~(3-4) (1992) 323--345.
\newblock

\bibitem{chen2009a}
Z.~Chen, M.~P. Burke, Y.~Ju,
  \href{https://ac.els-cdn.com/S154074890800285X/1-s2.0-S154074890800285X-main.pdf?_tid=9a918b04-e5bb-11e7-ba5d-00000aab0f6b&acdnat=1513798137_852f09aa532f445f0fc0e87185230472}{{Effects
  of Lewis number and ignition energy on the determination of laminar flame
  speed using propagating spherical flames}}, Proceedings of the Combustion
  Institute 32~(1) (2009) 1253--1260.
\newblock

\bibitem{Zhang2020e}
T.~Zhang, A.~J. Susa, R.~K. Hanson, Y.~Ju,
  \href{https://linkinghub.elsevier.com/retrieve/pii/S1540748920301449}{{Studies
  of the dynamics of autoignition assisted outwardly propagating spherical cool
  and double flames under shock-tube conditions}}, Proceedings of the
  Combustion Institute 000 (2020) 1--9.
\newblock

\bibitem{Ruetsch1995}
G.~R. Ruetsch, L.~Vervisch, A.~Li{\~{n}}{\'{a}}n,
  \href{http://aip.scitation.org/doi/10.1063/1.868531}{{Effects of heat release
  on triple flames}}, Physics of Fluids 7~(6) (1995) 1447--1454.
\newblock

\bibitem{Mizobuchi2002a}
Y.~Mizobuchi, S.~Tachibana, J.~Shinio, S.~Ogawa, T.~Takeno,
  \href{https://linkinghub.elsevier.com/retrieve/pii/S0082078404001948
  https://linkinghub.elsevier.com/retrieve/pii/S1540748902802450}{{A numerical
  analysis of the structure of a turbulent hydrogen jet lifted flame}},
  Proceedings of the Combustion Institute 29~(2) (2002) 2009--2015.
\newblock

\bibitem{Hernandez2018}
F.~E. Hern{\'{a}}ndez, N.~Mukhadiyev, X.~Xu, A.~Sow, B.~Jik, R.~Sankaran, H.~G.
  Im, {Direct numerical simulations of reacting flows with detailed chemistry
  using many-core / GPU acceleration R} 173 (2018) 73--79.
\newblock

\bibitem{Lu2021}
D.~Lu, H.~Wang, M.~Chen, L.~Lin, R.~Car, W.~E, W.~Jia, L.~Zhang,
  \href{https://doi.org/10.1016/j.cpc.2020.107624}{{86 PFLOPS Deep Potential
  Molecular Dynamics simulation of 100 million atoms with ab initio accuracy}},
  Computer Physics Communications 259 (2021) 107624.
\newblock \href {http://arxiv.org/abs/2004.11658} {\path{arXiv:2004.11658}},

\end{thebibliography}
\biboptions{sort&compress}

% -------------------------------------------------------------------- %
% -------------------------------------------------------------------- %
% -------------------------------------------------------------------- %

% -------------------------------------------------------------------- %
% -------------------------------------------------------------------- %
% -------------------------------------------------------------------- %

\end{document}